\documentclass[usenames,dvipsnames]{article}

\usepackage{a4wide}

\usepackage{amsmath,amsfonts,amsthm}
\usepackage[justification=justified, singlelinecheck=false]{caption}

\usepackage{tikz}

\usepackage{algorithm}
\usepackage{algpseudocode}

\usepackage{subfigure}
\usepackage{here} 
\usepackage[colorlinks=false, pdfborder={0 0 0}]{hyperref}
\usepackage{xpatch}
\usepackage[round]{natbib}

\setcitestyle{aysep={}}
\newcommand{\norm}[1]{\left\lVert#1\right\rVert}
\makeatletter
\g@addto@macro{\UrlBreaks}{\UrlOrds}
\makeatother

\begin{document}
	
\title{Spectral Density-Based and Measure-Preserving ABC \\ for partially observed diffusion processes \\ \begin{Large}\textit{An illustration on Hamiltonian SDEs}\end{Large}}
\author{Evelyn Buckwar, Massimiliano Tamborrino, Irene Tubikanec \\ Institute for Stochastics \\ Johannes Kepler University Linz, Austria}

\date{}
\maketitle	
	
\thispagestyle{empty}

\section*{Abstract}
Approximate Bayesian Computation (ABC) has become one of the major tools of likelihood-free statistical inference in complex mathematical models. Simultaneously, stochastic differential equations (SDEs) 
have developed to an established tool for modelling time dependent, real world phenomena with underlying random effects. When applying ABC to stochastic models, two major difficulties arise. First, the derivation of effective summary statistics and proper distances is particularly challenging, since simulations from the stochastic process under the same parameter configuration result in different trajectories. Second, exact simulation schemes to generate trajectories from the stochastic model are rarely available, requiring the derivation of suitable numerical methods for the synthetic data generation. 
To obtain summaries that are less sensitive to the intrinsic stochasticity of the model, 
we propose to build up the statistical method (e.g., the choice of the summary statistics) on the underlying structural properties of the model. Here, we focus on the existence of an invariant measure and we map the data to their estimated invariant density and invariant spectral density. Then, to ensure that these model properties are kept in the synthetic data generation, we adopt measure-preserving numerical splitting schemes. The derived property-based and mea\-sure-preserving ABC method is illustrated on the broad class of partially observed Hamiltonian {type} SDEs, both with simulated data and with real elec\-troen\-cephalo\-gra\-phy (EEG) data. The proposed ingredients can be incorporated into any type of ABC algorithm and directly  applied to all SDEs that are characterised by an invariant distribution and for which a measure-preserving numerical method can be derived.

\subsubsection*{Keywords} Approximate Bayesian Computation, Likelihood-free inference, 
Stochastic differential equations, Numerical splitting schemes, Invariant measure, Neural mass models

\subsubsection*{Acknowledgements} This research was partially supported by the Austrian Science Fund (FWF): W1214-N15, project DK14.

\section{Introduction}
\label{intro}
Over the last decades, SDEs have become an established and  powerful tool for modelling time dependent, real world phenomena with underlying random effects. They have been successfully applied to a variety of scientific fields, ranging from biology over finance, to physics, chemistry, neuroscience and others. Diffusion processes obtained as solutions of SDEs are typically characterised by some underlying structural properties whose investigation and preservation is crucial. Examples are boundary properties, symmetries or the preservation of invariants or qualitative behaviour such as the ergodicity or the conservation of energy. Here, we focus on a specific structural property, namely the existence of a unique invariant measure. Besides the modelling, it is of primary interest to estimate the underlying model parameters. This is particularly difficult when the multivariate stochastic process is only partially observed through a $1$-dimensional function of its coordinates (the output process), a scenario that we tackle here. Moreover, due to the increasing complexity of SDEs, needed to understand and reproduce the real data, the underlying likelihood is often unknown or intractable. Among several likelihood-free inference approaches, we focus on the simulation-based ABC method. We refer to \citet{25} and to the recently published book \lq\lq Handbook of Approximate Bayesian Computation\rq\rq \ for an exhaustive discussion \citep{sisson2018handbook}.

ABC has become one of the major tools for parameter inference in complex mathematical models in the last decade. The method is based on the idea of deriving an approximate posterior density targeting the true (unavailable) posterior by running massive simulations from the model to replace the intractable likelihood. It was first introduced in the context of population genetics; see, e.g., \cite{34}. Since then, it has been successfully applied in a wide range of fields; see, e.g., \cite{37,21,Boysetal2008,35,Mooresetal2015,51}. Moreover, ABC has also been proposed to infer parameters from time series models \citep[see, e.g.,][]{Drovandietal,41}, state space models \citep[see, e.g.,][]{Martinetal2018,Tancredi} and SDE models \citep[see, e.g.,][]{52,Maybank2017,43,44,53,46,47}. Several advanced ABC algorithms have been proposed in the literature, such as, ABC-SMC, ABC-MCMC, sequential-annealing ABC, noisy ABC; see, e.g., \cite{FanSisson2018} and the references there\-in for a recent review. The idea of the basic acceptance-rejection algorithm is to keep a sampled parameter value from the prior as a realisation from the approximate posterior, if the distance between the summary statistics of the synthetic dataset, which is generated conditioned on this parameter value, and the summaries of the original reference data is smaller than some tolerance level. The goal of this paper is to illustrate how building up the ABC method on the {structural properties} of the underlying SDE and using a numerical method capable to preserve {them} in the generation of the data from the model leads to a successful inference even when applying ABC in this basic acceptance-rejection form. 

The performance of any ABC method depends heavily on the choice of \lq\lq informative enough\rq\rq\ summary statistics, a suitable distance measure and a proper tolerance level $\epsilon$. The quality of the approximation improves as $\epsilon$ decreases, and it has been shown that, under some conditions, the approximated ABC posterior converges to the true one when $\epsilon\to 0$ \citep{41}. At the same time though, the computational cost increases when $\epsilon$ decreases. A possibility is to use ad-hoc threshold selection procedures; see, e.g., \cite{22,33,12,28,26}. Here, we fix the tolerance level $\epsilon$ as a percentile of the calculated distances. This is another common practice used, for example, in \cite{34,20,46,50}. Instructions for constructing effective summaries and distances are rare and they depend on the problem under consideration; see, e.g.,  \cite{38} for  a semi-automatic linear regression approach, \cite{Jiang2017} for an automatic construction approach based on training deep neural networks and \cite{33,32} for two recent reviews. To avoid the information loss caused by using non-sufficient summary statistics another common procedure is to work with the entire dataset; see, e.g., \cite{41,46}. This requires the application of more sophisticated distances $d$ such as the Wasserstein metric \citep{10,11} or other distances designed for time series; for an overview see, e.g., \cite{REF_I4} and the references therein.

When working with stochastic models,  simulations from the stochastic simulator, conditionally to the same parameter  configuration, yield different trajectories. To consider summary statistics that are less sensitive to the intrinsic stochasticity of the model \citep{Wood2010}, we choose
them based on the structural property of an underlying invariant measure. The idea is to map the data, i.e., the realisations of the output process, to an object that is invariant for repeated simulations under the same parameter setting and that reacts sensitive to small changes in the parameters. In particular, 
we map the data to their estimated invariant density and invariant spectral density, taking thus the dependence structure of the dynamical model into account. The distance measure can then be chosen according to the mapped data.

As other simulation-based statistical methods, e.g., MCMC, SMC or machine learning algorithms, ABC relies on the ability of simulating data from the model. However, the exact simulation from complex stochastic models is rarely possible, and thus numerical methods need to be applied. This introduces a new level of approximation into the ABC framework. When the statistical method is build upon the structural properties of the underlying model, the successful inference can only be guaranteed when these properties are preserved in the synthetic data generated from the model. However, the issue of deriving a property-preserving numerical method when applying ABC to SDEs is usually seen as not {so} relevant, and it is 
usually recommended to use the {Euler-Maruya\-ma} scheme or one of the higher order approximation methods described in \cite{REF_I3}; see, e.g., \cite{43,44,53,46}. In general, these standard methods do not preserve the underlying structural properties of the model; see, e.g., \cite{2,REF13_1,REF13_2,57}.

Here, we propose to apply structure-preserving numerical splitting schemes within the ABC algorithm. The idea of {these} methods is to split the SDE into explicitly solvable subequations and to apply a proper composition of the resulting exact solutions. Standard procedures are, for {example}, the Lie-Trotter method and the usually more accurate Strang approach; see, e.g., \cite{4}. Since the only approximation enters through the composition of the derived explicit solutions, numerical splitting schemes usually preserve the structural properties of the underlying SDE and accurately reproduce its qualitative behaviour.  Moreover, {they} usually have the same order of convergence as the frequently applied Euler-Maruyama method and are likewise efficient. We refer to \cite{Blanes2009} and \cite{Mclachlan2002} for an exhaustive discussion of splitting methods for broad classes of ordinary differential equations (ODEs), which partially have already been carried over to SDEs; see, e.g., \cite{REF_I7} for a general class of SDEs, \cite{Ableidinger2016} for the stochastic Landau-Lifshitz equations, \cite{Brehier2018} for the Allen-Cahn equation and \cite{2} for Hamiltonian type SDEs.

The main contribution of this work lies in the combination of the proposed invariant measure-based summary statistics and the measure-preserving numerical splitting schemes within the ABC framework. We demonstrate that a simulation-based inference method, here ABC, can only perform well if the underlying simulation method preserves the structural properties of the SDE. While the use of preserving splitting schemes within the ABC method yield successful results, applying a general purpose numerical method, such as the Euler-Maruyama discretisation, may result in seriously wrong inferences. 
We illustrate the proposed Spectral Density-Based and Measure-Preserving ABC method on the class of stochastic Hamiltonian type equations for which the existence of an underlying unique invariant distribution and measure-preserving numerical splitting schemes have been already intensively studied in the literature; see, e.g., \cite{2,3,6,5}. Hamiltonian type SDEs have been investigated in molecular dynamics, where they are typically referred to as Langevin equations; see, e.g., \cite{6}. Recently, they have also received considerable attention in the field of neuroscience as the so-called neural mass models \citep{2}.

The paper is organised as follows. In Section \ref{sec:2}, we recall the acceptance-rejection ABC setting. We introduce the invariant measure-based summary statistics and propose a proper distance. We then discuss the importance of considering measure-preserving numerical schemes for the synthetic data generation when exact simulation methods are not applicable and provide a short introduction to numerical splitting methods. In Section \ref{sec:3}, we introduce Hamiltonian type SDEs and recall two splitting integrators preserving the invariant measure of the model. In Section \ref{sec:4}, we validate the proposed method by investigating the stochastic harmonic oscillator, for which exact simulation is possible. In Section \ref{sec:5}, we apply the proposed ABC method to the stochastic Jansen and Rit neural mass model (JR-NMM). We refer to \cite{1} for the original version, an ODE with a stochastic input function, and to \cite{2} for its reformulation as a Hamiltonian type SDE. This model has been reported to successfully reproduce  EEG data. We illustrate the performance of the proposed ABC method with both simulated and real data. Final remarks, possible extensions and conclusions are reported in Section \ref{sec:6}. Further illustrations of the proposed ABC method are available in the provided supplementary material, here reported as Section \ref{sec:7}. A sample code used to generate the main results is available on github.

\section{Spectral Density-Based and Measure-Preserving ABC for partially observed SDEs with an invariant distribution}
\label{sec:2}
Let $(\Omega,\mathcal{F},\mathbb{P})$ be a complete probability space with the right-continuous and complete filtration $\mathbb{F}=\{\mathcal{F}\}_{t \in [0,T]}$. Let  $\theta=(\theta_1,...,\theta_k)$, $k \in \mathbb{N}$, be a vector of relevant model parameters.
We consider the following $n$-dimensional, $n \in \mathbb{N}$, non-autonomous SDE of It\^{o}-type describing the time evolution of a system of interest
\begin{align}
	\begin{split}
		dX(t) &=f(t,X(t);\theta) \ dt + \mathcal{G} (t, X(t);\theta) \ dW(t) \\
		X(0) &= X_0, \ t \in [0,T].
	\end{split}
	\label{SDE}
\end{align}
The initial value $X_0$ is either deterministic or a $\mathbb{R}^n$-valued random variable, measurable with respect to $\mathbb{F}$. Here, $\textbf{W}=(W(t))_{t \in [0,T]}$ is a $r$-dimensional, $r \in \mathbb{N}$, Wiener process with independent and $\mathbb{F}$-adapted components. We further assume that the drift component $f:~[0,T] \times \mathbb{R}^n \to \mathbb{R}^n$ and the diffusion component $\mathcal{G}: [0,T] \times \mathbb{R}^n \to \mathbb{R}^{n \times r}$ fulfil the necessary global Lipschitz and linear growth conditions, such that the existence and the pathwise uniqueness of an $\mathbb{F}$-adapted strong solution process $\textbf{X}=(X(t))_{t \in [0,T]} \in \mathbb{R}^n$ of \eqref{SDE} is guaranteed; see, e.g., \cite{9}.  

We aim to infer the parameter vector $\theta$ inherent in the SDE \eqref{SDE}, when the $n$-dimensional solution process  $\textbf{X}$ is only partially observed through the 1-dimensional and parameter-dependent output process
\begin{equation}
	\textbf{Y}_\theta=(Y_\theta(t))_{t \in [0,T]}=g(\textbf{X}),
	\label{output}
\end{equation}
where  $g: \mathbb{R}^n \to \mathbb{R}$ is a real-valued continuous function of the components of $\textbf{X}$.

Further, we assume a specific underlying structural model property, namely the existence of a unique invariant measure $\eta_{\textbf{Y}_\theta}$ on $(\mathbb{R},\mathcal{B}(\mathbb{R}))$ of the output process $\textbf{Y}_\theta$, where $\mathcal{B}$ denotes the Borel Sigma-algebra.
The process has invariant density $f_{\textbf{Y}_\theta}$ and mean, autocovariance and variance  given by
\begin{align} \label{Stationary}
	\begin{split}
		\mathbb{E}[Y_\theta(t)] &= \eta_{\mu} \in \mathbb{R}, \\
		\textrm{Cov}[Y_\theta(t),Y_\theta(s)]&:=r_{\theta}(t,s)=r_{\theta}(t-s), \ s \leq t,
		\\
		\textrm{Var}[Y_\theta(t)]&=r_{\theta}(0)=\eta_{\sigma^2} \in \mathbb{R}^+.
	\end{split}
\end{align}
If the solution process $\textbf{X}$ of SDE \eqref{SDE} admits an invariant distribution $\eta_{\textbf{X}}$ on $(\mathbb{R}^n,\mathcal{B}(\mathbb{R}^n))$, then the output process $\textbf{Y}_\theta$ inherits this structural property by means of the marginal invariant distributions of $\eta_{\textbf{X}}$. Furthermore, if $X(0) \sim \eta_{\textbf{X}}$, then the process $\textbf{Y}_\theta=(Y_\theta(t))_{t \in [0,\infty)}$ evolves according to the distribution $\eta_{\textbf{Y}_{\theta}}$ for all $t \geq 0$.
Our goal is to perform statistical inference for the parameter vector $\theta$ of the SDE \eqref{SDE}, when the solution process \textbf{X} is partially observed through discrete time measurements of the output  process $\textbf{Y}_\theta$ given in \eqref{output}, by benefiting from the (in general unknown) invariant distribution $\eta_{\textbf{Y}_\theta}$ satisfying \eqref{Stationary}.

\subsection{The ABC method}
\label{subsec:2:1}
Let $y=(y(t_i))_{i=1}^l$, $l \in \mathbb{N}$, be the reference data, corresponding to discrete time observations of the output process $\textbf{Y}_\theta$. Let us denote by $\pi(\theta)$ and $\pi(\theta|y)$ the prior and the posterior density, respectively. For multivariate complex SDEs, the underlying likelihood is often unknown or intractable. The idea of the ABC method is to derive an approximate posterior density for $\theta$ by replacing the unknown likelihood by possibly billions of synthetic dataset simulations generated from the underlying model \eqref{SDE} and mapped to ${\bf Y}_\theta$ through \eqref{output}. The basic acceptance-rejection ABC algorithm consists of three steps: i. Sample a value $\theta'$ from the prior $\pi(\theta)$; ii. Conditionally on $\theta'$, simulate a new artificial dataset from the model \eqref{SDE} and derive the synthetic data $y_{\theta'}=(y_{\theta'}(t_i))_{i=0}^m,  t_0=0,t_m=T,m\in\mathbb{N}$, from the process $\bf Y_{\theta'}$ given by \eqref{output}; iii. Keep the sampled parameter value $\theta'$ as a realisation from the posterior if the distance $d$ between a vector of summary statistics $s=(s_1,\ldots, s_h), h \in \mathbb{N}$, of the original and the synthetic data is smaller than some threshold level $\epsilon\geq 0$, i.e., $d(s(y),s(y_{\theta'}))<\epsilon$. 
\begin{algorithm}
	\caption{Acceptance-rejection ABC
		\ \\ \textbf{Input:} Observed data $y$ \ \\
		\textbf{Output:} Samples from the posterior $\pi_{\textrm{ABC}} (\theta|y)$	
	}\label{Algorithm_Standard}
	\begin{algorithmic}[1]
		\State Precompute a vector of summary statistics $s(y)$
		\State Choose a prior distribution $\pi(\theta)$ and a tolerance level $\epsilon$
		\For{$i = 1:N$}
		\State Draw $\theta^i=(\theta_1^i,...,\theta_k^i)$ from the prior $\pi(\theta)$
		\State Conditionally on $\theta^i$, simulate a new realisation $y_{\theta^i}$ from the output process $\textbf{Y}_\theta$
		\State Compute the summaries $s(y_{\theta^i})$
		\State Calculate the distance $D_i=d(s(y),s(y_{\theta^i}))$
		\State If $D_i<\epsilon$, keep $\theta^i$ as a sample from the posterior
		\EndFor
	\end{algorithmic}
\end{algorithm}

When $\epsilon=0$ and $s$ is a vector of sufficient statistics for $\theta$, the acceptance-rejection ABC (summarised in Algorithm \ref{Algorithm_Standard}) produces samples from the true posterior $\pi(\theta|y)$.
Due to the complexity of the underlying SDE \eqref{SDE}, we cannot derive non-trivial sufficient statistics $s$ for $\theta$.
Moreover, due to the underlying stochasticity of the model, $\mathbb{P}(d(s(y),s(y_{\theta'}))=0)=0$. Thus, $\epsilon$ is required to be strictly positive. Hence, the acceptance-rejection ABC Algorithm \ref{Algorithm_Standard} yields samples from an approximated posterior $\pi_{\textrm{ABC}} (\theta|y)$ according to
\begin{equation*}
	\pi(\theta|y) \approx \pi_{\textrm{ABC}} (\theta|y) = \pi\{ \theta  |  d(s(y),s(y_\theta))<\epsilon\}.
\end{equation*}

Besides the tolerance level $\epsilon$, which we fix as a percentile of the calculated distances, the quality of the ABC method depends strongly on the choice of suitable summary statistics combined with a proper distance measure and on the numerical method used to generate the synthetic data from the model.
In the following, we introduce summaries that are very effective for the class of models having an underlying invariant distribution, we suggest a proper distance based on them and we propose the use of measure-preserving numerical splitting schemes.

\subsection{An effective choice of summaries and distances: Spectral Density-Based ABC}
\label{subsec:2:2}
\begin{figure*}
	\includegraphics[width=1.0\textwidth]{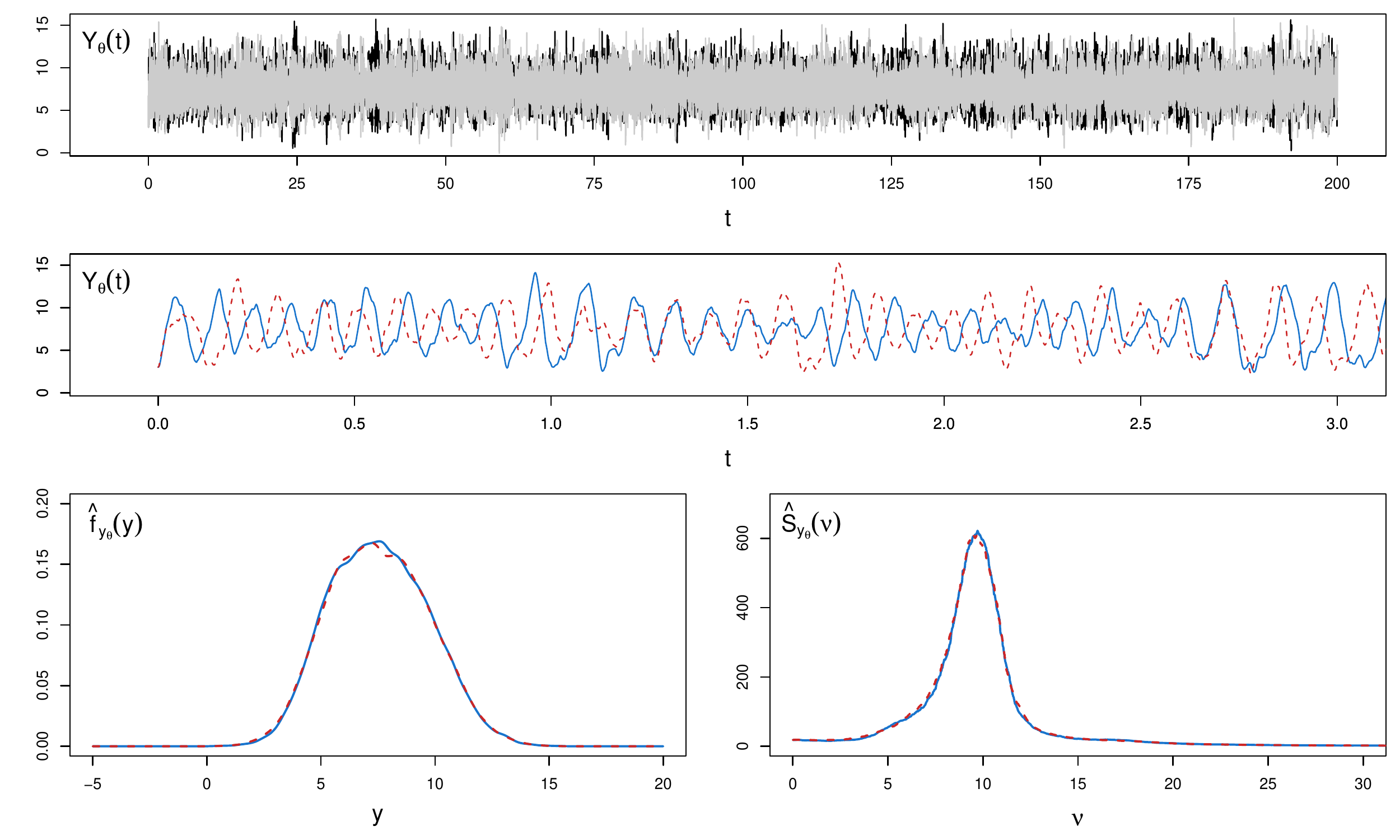}\\
	\caption{Two realizations of the output process of the stochastic JR-NMM \eqref{JR-NMM} generated with the numerical splitting method \eqref{Strang} for an identical choice of $\theta$. The lengths of the time intervals are $T=200$ and $T=3$ (to provide a zoom) in the top and middle panel, respectively.  The two invariant densities and two invariant spectral densities,  estimated from the two full datasets shown in the top panel, are reported in the lower panel on the left and right, respectively}
	\label{2P}
\end{figure*}

When applying ABC to stochastic models, an important statistical challenge arises. Due to the intrinsic randomness, repeated simulations of the process $\textbf{Y}_\theta$ under the same parameter vector $\theta$ may yield very different trajectories. An illustration is given in Figure \ref{2P} (top and middle panels), where we report two trajectories of the output process of the stochastic JR-NMM \eqref{JR-NMM} generated with an identical parameter configuration. This model is a specific SDE of type \eqref{SDE}, observed through $\textbf{Y}_\theta$ as in \eqref{output}, and admitting an invariant distribution $\eta_{\textbf{Y}_{\theta}}$ satisfying \eqref{Stationary}. See Section \ref{sec:5} for a description of the model. In the top panel, we visualise the full paths for a time $T=200$, while in the middle panel we provide a zoom, showing only the initial part. 

\medskip

\noindent{\bf{Proposal 1: To use the property of an invariant measure $\eta_{\bf Y_\theta}$ and to map the data $y_\theta$ to their estimated invariant density $\hat{f}_{y_\theta}$ and invariant spectral density $\hat{S}_{y_\theta}$.}}\smallskip

Instead of working with the output process $\textbf{Y}_\theta$, we take advantage of the structural model property $\eta_{\textbf{Y}_{\theta}}$ and focus on its invariant density $f_{\textbf{Y}_\theta}$ and its invariant spectral density $S_{\textbf{Y}_\theta}$. Both are deterministic functions characterized by the underlying parameters $\theta$, and thus invariant for repeated simulations under the same parameter configuration. The invariant spectral density is obtained from the Fourier transformation of the autocovariance function $r_{\theta}$, and it is given by
\begin{equation} \label{Spec}
	S_{\textbf{Y}_\theta}=\mathcal{F}\{ r_\theta \} (\omega)=\int_{-\infty}^{\infty} r_{\theta}(\tau)e^{-i\omega \tau} \ d\tau,
\end{equation}
for $\omega \in [-\pi,\pi]$. The angular frequency $\omega$ relates to the ordinary frequency $\nu$ via $\omega=2\pi \nu$. Since both $f_{\textbf{Y}_\theta}$ and $S_{\textbf{Y}_\theta}$ are typically unknown, we estimate them from a dataset $y_\theta$.  
First, we estimate the invariant density $f_{\textbf{Y}_\theta}$ with a kernel density estimator, denoted by $\hat{f}_{y_\theta}$; see, e.g., \cite{REF_I2}. Second, we estimate the invariant spectral density $S_{\textbf{Y}_\theta}$ \eqref{Spec} with a smoothed periodogram estimator \citep{61,REF_I5}, denoted by $\hat{S}_{y_\theta}$, which is typically evaluated at Fourier frequencies. 
Differently from the invariant density, the invariant spectral density does not account for the mean $\mathbb{E}[\textbf{Y}_\theta]$ but captures the dependence structure of the data coming from the model. We define the invariant measure-based summary statistics $s$ of a dataset $y_\theta$ as 
\begin{equation}\label{choice_s}
	s(y_\theta):=(\hat{S}_{y_\theta},\hat{f}_{y_\theta}).
\end{equation} 
Figure \ref{2P} shows the two estimated invariant densities (left lower panel) and invariant spectral densities (right lower panel), all derived from the full paths of the output process $\textbf{Y}_\theta$ (top panel).

After performing the data mapping \eqref{choice_s}, which significantly reduces the randomness in the output of the stochastic simulator, the distance $d$ can be chosen among the distance measures between two $\mathbb{R}$-valued functions. Here, we consider the integrated absolute error (IAE) defined by
\begin{equation}\label{RIAE}
	\text{IAE}(g_1,g_2):=\int\limits_{\mathbb{R}} \ \Bigl|g_1(x)-g_2(x)\Bigr| \ dx \in\mathbb{R}^+.
\end{equation}
Another natural possibility could be a distance chosen among the so-called f-divergences \citep[see, e.g.,][]{REF_I6}, or the Wasserstein distance, recently proposed for ABC \citep{10}. Within the ABC framework (see Step $7$ in Algorithm \ref{Algorithm_Standard}), we suggest to use the following distance
\begin{equation}\label{weight}
	d(s(y),s(y_\theta)):= \text{IAE}(\hat{S}_{y},\hat{S}_{y_\theta})+w \cdot \text{IAE}(\hat{f}_{y},\hat{f}_{y_\theta}),
\end{equation}
returning a weighted sum of the areas between the densities estimated from the original and the synthetic data\-sets. Here, $w \geq 0$ is a weight that we assign to the part related to the IAE of the invariant densities such that the two errors are of the same \lq\lq order of magnitude\rq\rq. This is particularly needed because, differently from the invariant density, the invariant spectral density does not integrate to 1. We obtain a value for the weight by performing an ABC pilot simulation. It consists in reiterating the following steps $L$ times:
\begin{algorithmic}[1]	
	\State Draw $\theta'$ from the prior $\pi(\theta)$
	\State Conditionally on $\theta'$, simulate two artificial datasets	\newline
	$y_{\theta'}^1$ and $y_{\theta'}^2$ from the output process $\textbf{Y}_\theta$
	\State Compute the corresponding summaries \eqref{choice_s}, i.e.,  $s(y_{\theta'}^1)=(\hat{S}_{y_{\theta'}^1},\hat{f}_{y_{\theta'}^1})$ and $s(y_{\theta'}^2)=(\hat{S}_{y_{\theta'}^2},\hat{f}_{y_{\theta'}^2})$
	\State Determine a value for the weight using \eqref{weight}, i.e.,  $w'=\frac{\text{IAE}(\hat{S}_{y_{\theta'}^1},\hat{S}_{y_{\theta'}^2})}{\text{IAE}(\hat{f}_{y_{\theta'}^1},\hat{f}_{y_{\theta'}^2})}$ 
\end{algorithmic} 
\ \\
Then, we take the median of the resulting $L$ values $w'$. See, e.g., \cite{Prangle2017} for alternative approaches for the derivation of weights among summary statistics.
Since the densities $\hat f_{{y_\theta}}$ and $\hat S_{y_\theta}$ are estimated at discrete points, the IAE \eqref{RIAE} is approximated applying trapezoidal integration. 

In Algorithm \ref{Algorithm_Standard}, we 
assume to observe $M \in \mathbb{N}$ data\-sets 
referring to $M$ realisations of the output process $\textbf{Y}_\theta$ sampled at $l \in \mathbb{N}$ discrete points in time, resulting in a matrix $y \in \mathbb{R}^{M \times l}$ of observed data. 
The median of the distances \eqref{weight} computed for each of the $M$ datasets 
\begin{equation}\label{7b}
	D=
	\textrm{median}\left\{ \left( 
	\textrm{IAE}(\hat{S}_{y_k},\hat{S}_{{y}_{\theta}}) + w \cdot
	\textrm{IAE}(\hat{f}_{y_k},\hat{f}_{{y}_{\theta}}) \right)_{k=1}^M \right\}
\end{equation}\noindent
is then returned as a global distance in Step 7. Other strategies can be adopted.  For example, considering the mean instead yields similar results in all our experiments. One can interpret $y$ as a long-time trajectory (when using simulated observed reference data) or as a long-time recording of the modelled phenomenon (when using real observed reference data) that is cut into $M$ pieces. Alternatively, $y$ would consist of M independent repeated experiments or simulations, when  dealing with real or simulated data, respectively. As expected, having $M>1$ datasets improves the quality of the estimation due to the increased number of observations.

\subsection{A new proposal of synthetic data generation: Measure-Preserving ABC}
\label{subsec:2:3}
\begin{figure*}
	\includegraphics[width=1.0\textwidth]{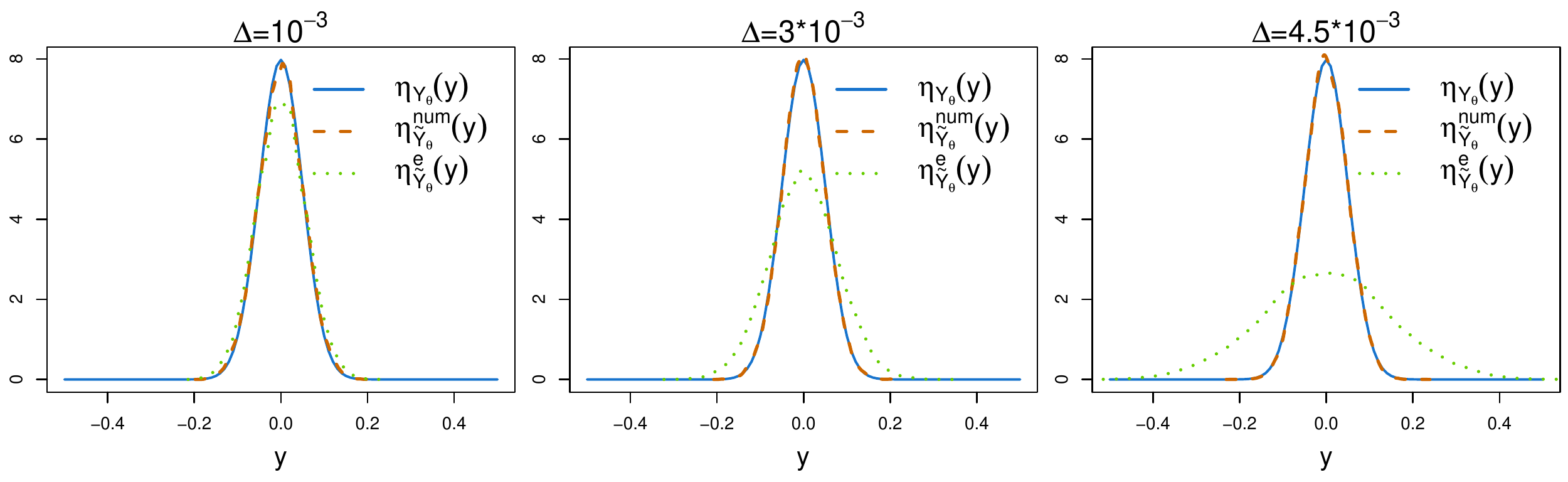}\\
	\caption{Comparison of the true invariant density of the weakly damped stochastic harmonic oscillator \eqref{MP} (blue solid lines) with the densities estimated using a kernel density estimator applied on data $y_\theta$ generated by the measure-preserving splitting scheme \eqref{Strang_2} (orange dashed lines) and the Euler-Maruyama method \eqref{EM} (green dotted lines) with time step $\Delta$ up to time $T=10^3$. The values of the time steps are $\Delta=10^{-3}$ (left figure), $3\cdot 10^{-3}$ (central figure) and $4.5 \cdot 10^{-3}$ (right figure), respectively}
	\label{Plot_inv}
\end{figure*}
A crucial aspect of ABC and of all other simulation-based methods is the ability of simulating from the model (Step $5$ of Algorithm \ref{Algorithm_Standard}).  
Consider a discretized time grid with the equidistant time step $\Delta=t_{i+1}-t_{i}$ and
let $\tilde{y}_\theta=(\tilde{y}_\theta(t_i))_{i=1}^m$ be a realisation from the   process ${\bf \widetilde Y_\theta}=(\widetilde Y_\theta(t_i))_{i=1}^m$, obtained through a numerical method, approximating $\textbf{Y}_\theta$ at the discrete data points, i.e., $\widetilde{Y}_\theta(t_i) \approx Y_\theta(t_i)$. The lack of exact simulation schemes, i.e., $\widetilde{Y}_\theta(t_i) = Y_\theta(t_i)$, introduces a new level of approximation in the statistical inference. In particular, Algorithm \ref{Algorithm_Standard} samples from an approximated posterior density of the form
\begin{equation*}
	\pi(\theta|y) \approx \pi^{\textrm{num}}_{\textrm{ABC}} (\theta|y) := \pi\{ \theta  |  d(s(y),s(\tilde y_\theta))<\epsilon\}.
\end{equation*}
As a consequence, $y_\theta$ in Step $5$ of Algorithm \ref{Algorithm_Standard} is replaced by its numerical approximation $\tilde{y}_\theta$.

The commonly used Euler-Maruyama scheme yields discretised trajectories of the solution process $\textbf{X}$ of the SDE \eqref{SDE} through \citep{REF_I3}
\begin{equation}\label{EM}
	\widetilde{X}(t_{i+1})=\widetilde{X}(t_i)+f(t_i,\widetilde{X}(t_i);\theta) \Delta + \mathcal{G}(t_i,\widetilde{X}(t_i);\theta) \xi_i, 
\end{equation}
where $\xi_i$ are Gaussian vectors with null mean and variance $\Delta \mathbb{I}_n$, where $\mathbb{I}_n$ denotes the $n \times n$-dimensional identity matrix. As previously discussed, in general, the Euler-Maruyama method does not preserve the underlying invariant distribution $\eta_{\bf Y_\theta}$.

\medskip 

\noindent{\bf{Proposal 2}: To adopt a numerical method for the synthetic data generation that preserves the underlying invariant measure of the model.}\smallskip

We apply numerical splitting schemes within the ABC framework and provide a brief account of their theory. Let us assume that the drift $f$ and the diffusion $\mathcal{G}$ of SDE \eqref{SDE} can be written as
\begin{equation*}
	f(t,X(t);\theta)=\sum_{j=1}^{d} f^{[j]}(t,X(t);\theta), \quad 
	\mathcal{G}(t,X(t);\theta)=\sum_{j=1}^{d} \mathcal{G}^{[j]}(t,X(t);\theta), \quad d \in \mathbb{N}.
\end{equation*}
The goal is to decompose $f$ and $\mathcal{G}$ in a way such that the resulting $d$ subequations
\begin{equation*}
	dX(t)=f^{[j]}(t,X(t);\theta) \ dt + \mathcal{G}^{[j]} (t, X(t);\theta) \ dW(t), 
\end{equation*}
for $j \in \{ 1,\dots,d \}$, can be solved exactly. Note that, the terms $\mathcal{G}^{[j]}$ can be null, resulting in deterministic equations (ODEs). Let $X^{[j]}(t)=\varphi_t^{[j]}(X_0)$ denote the exact solutions (flows) of the above subequations at time $t$ and starting from $X_0$. Once these explicit solutions are derived, a proper composition needs to be applied.  
Here we use the Strang approach
\begin{equation*}
	\left( \varphi_{\Delta/2}^{[1]} \circ ... \circ \varphi_{\Delta/2}^{[d-1]} \circ \varphi_{\Delta}^{[d]} \circ \varphi_{\Delta/2}^{[d-1]} \circ ... \circ \varphi_{\Delta/2}^{[1]} \right)(x), \quad x \in \mathbb{R}^n,
\end{equation*}
that provides a numerical solution for the original SDE \eqref{SDE}.

In Figure \ref{Plot_inv}, we illustrate how the numerical splitting method preserves the underlying invariant measure of the weakly damped stochastic harmonic oscillator \eqref{MP}, independently from the choice of the time step $\Delta$. This is a specific SDE of type \eqref{SDE}, observed through $\textbf{Y}_\theta$ as in \eqref{output} and with a known invariant distribution $\eta_{\textbf{Y}_{\theta}}$. See Section \ref{sec:3} for the detailed numerical splitting scheme and Section \ref{sec:4} for a description of the model. In contrast, the Euler-Maruyama scheme performs worse as $\Delta$ increases. Each subplot shows a comparison of the true invariant density (blue solid lines) and the corresponding kernel estimate $\hat{f}_{y_\theta}$ based on a  path $y_\theta$ from the model, generated from the measure-preserving numerical splitting scheme \eqref{Strang_2} (dashed orange lines) or the Euler-Maruyama approach (dotted green lines). The data are generated under $T=10^3$ and different values for the time step, namely $\Delta=10^{-3}$, $3\cdot10^{-3}$, $4.5\cdot10^{-3}$. 

\subsection{Notation}
\label{subsubsec:2:4}
We apply the summary statistics \eqref{choice_s} and the distance \eqref{7b} in Algorithm \ref{Algorithm_Standard}. We use the notation Algorithm \ref{Algorithm_Standard} (i) for the Spectral Density-Based ABC method when the synthetic data are simulated exactly, Algorithm \ref{Algorithm_Standard} (ii) for the Spectral Density-Based and Measure-Preserving ABC method when a measure-preserving numerical splitting scheme is applied and Algorithm \eqref{Algorithm_Standard} (iii) when we generate the data with the non-preserving Euler-Maruyama scheme.

To evaluate the performance of the proposed ABC method, we analyse the marginal posterior densities, denoted by $\pi_{\textrm{ABC}}^*(\theta_j|y)$, $j \in \{1,...,k\}$, obtained from the posterior density $\pi_{\textrm{ABC}}^*(\theta|y)$ corresponding to $\pi_{\textrm{ABC}}(\theta|y)$, $\pi_{\textrm{ABC}}^\textrm{num}(\theta|y)$ or $\pi_{\textrm{ABC}}^e(\theta|y)$, depending on whether we obtain it from Algorithm \ref{Algorithm_Standard} (i), (ii) or (iii).
Following this notation, we define by $\hat{\theta}_{\textrm{ABC},j}^*$ the marginal ABC posterior means.

\section{An illustration on Hamiltonian type SDEs}
\label{sec:3}
We illustrate the proposed ABC approach on Hamiltonian type SDEs and define the $n$-dimensional ($n=2d$, $d\in\mathbb{N}$) stochastic process
\begin{equation*}
	\textbf{X}:=(\textbf{Q},\textbf{P})^{'}=(Q(t),P(t))^{'}_{t \in [0,T]},
\end{equation*}
consisting of the two $d$-dimensional components
\begin{equation*}
	\textbf{Q}=(\mathbf{X_1},...,\mathbf{X_d})^{'} \ \text{and} \ \textbf{P}=(\mathbf{X_{d+1}},...,\mathbf{X_{2d}})^{'},
\end{equation*}
where $^{'}$ denotes the transpose.
The $n$-dimensional SDE of Hamiltonian type with initial value $X_0=~(Q_0,P_0)^{'}$ and $d$-dimensional ($r=d$) Wiener process $\textbf{W}$ describes the time evolution of the process $\textbf{X}$ by
\begin{equation}\label{Hamiltonian}
d \underbrace{
	\begin{pmatrix}
	Q(t) \\
	P(t) 
	\end{pmatrix}}_{X(t)} 
= 
\underbrace{
	\begin{pmatrix}
	\nabla_P H(Q(t),P(t)) \\
	-\nabla_Q H(Q(t),P(t)) -2\Gamma_\theta P(t)+G(Q(t);\theta) 
	\end{pmatrix}}_{f(X(t);\theta)} dt+
\underbrace{
	\begin{pmatrix}
	\mathbb{O}_d \\
	\Sigma_\theta
	\end{pmatrix}}_{\mathcal{G}(\theta)} dW(t).
\end{equation}
We denote with $\mathbb{O}_d$ the $d \times d$-dimensional zero matrix and with $\nabla_Q$ and $\nabla_P$ the gradient with respect to $Q$ and $P$, respectively. The SDE \eqref{Hamiltonian} consists of $4$ parts, each representing a specific type of behaviour. In this configuration, the first is the \textit{Hamiltonian part} involving $H:\mathbb{R}^d \times \mathbb{R}^d \to \mathbb{R}_0^+$ given by
\begin{equation*}
	H(\textbf{Q},\textbf{P}):=\frac{1}{2}(\norm{\textbf{P}}^2_{\mathbb{R}^d}+\norm{\Lambda_\theta \textbf{Q}}^2_{\mathbb{R}^d}),
\end{equation*}
where $\Lambda_\theta=\text{diag}[\lambda_{1},...,\lambda_{d}] \in \mathbb{R}^{d \times d}$ is a diagonal matrix. The second is the \textit{linear damping part}, described by the matrix $\Gamma_\theta=\text{diag}[\gamma_{1},...,\gamma_{d}] \in \mathbb{R}^{d \times d}$. The third is the \textit{non-linear displacement part}, consisting of the non-linear and globally Lipschitz continuous function $G: \mathbb{R}^d \to \mathbb{R}^d$. The fourth corresponds to the \textit{diffusion part}, given by $\Sigma_\theta=\text{diag}[\sigma_{1},...,\sigma_{d}] \in \mathbb{R}^{d \times d}$. 

\subsection{Structural model property}
\label{subsec:3:1}
Under the requirement of non-degenerate matrices $\Lambda_\theta$, $\Gamma_\theta$ and $\Sigma_\theta$, i.e., strictly positive diagonal entries, Hamiltonian type SDEs as in \eqref{Hamiltonian} are {often} ergodic.  
As a consequence, the distribution of the solution process $\textbf{X}$ (and thus of the output process ${\bf Y}_\theta$) converges exponentially fast towards a unique invariant measure $\eta_{\textbf{X}}$ on $(\mathbb{R}^n,\mathcal{B}(\mathbb{R}^n))$ (and thus $\eta_{\bf Y_\theta}$ on $(\mathbb{R},\mathcal{B}(\mathbb{R}))$; see, e.g., \cite{2} and the references therein.

\subsection{Measure-Preserving numerical splitting schemes}
\label{subsec:3:2}
Two splitting approaches for SDE \eqref{Hamiltonian} are provided, see \cite{2}.
Due to the non-linear term $G$, the SDE \eqref{Hamiltonian} cannot be solved explicitly.
With the purpose of excluding $G$, the Hamiltonian {type} SDE \eqref{Hamiltonian} is split into the two subsystems
\begin{equation}\label{Sub1}
d\begin{pmatrix} Q(t)\\ P(t) \end{pmatrix}=\underbrace{\begin{pmatrix} \nabla_P H((t),P(t)) \\ -\nabla_Q H(Q(t),P(t)) - 2 \Gamma_\theta P(t) \end{pmatrix}}_{f^{[1]}(X(t);\theta)}dt + \underbrace{
	\left( \begin{array}{c}\mathbb{O}_d\\\Sigma_\theta \end{array} \right)}_{\mathcal{G}^{[1]}(\theta)} dW(t),
\end{equation}
\begin{equation} \label{Sub2}
	d\begin{pmatrix} Q(t) \\ P(t)  \end{pmatrix}=\underbrace{\begin{pmatrix} 0_d \\ G(Q(t);\theta) \end{pmatrix}}_{f^{[2]}(Q(t);\theta)}dt,
\end{equation}
where $0_d$ denotes the $d$-dimensional zero vector.
This results in the linear SDE with additive noise \eqref{Sub1} and the non-linear ODE \eqref{Sub2} that can be both explicitly solved. Indeed, since $\nabla_P H(Q(t),P(t))=P(t)$ and $\nabla_Q H(Q(t),P(t))=\Lambda_\theta^2 Q(t)$, Subsystem \eqref{Sub1} can be rewritten as
\begin{equation} \label{narrow_SDE}
	dX(t)=A \cdot X(t) \ dt+B \ dW(t), \quad t \geq 0,
\end{equation}
with $A=
\begin{pmatrix}
\mathbb{O}_d & \mathbb{I}_d\\
-\Lambda_\theta^2 & -2\Gamma_\theta
\end{pmatrix}
$ and $B=\left( \begin{array}{c}\mathbb{O}_d\\\Sigma_\theta \end{array} \right)$.  
The exact path of System \eqref{narrow_SDE} is obtained through 
\begin{equation} \label{Exact_SDE}
	X(t_{i+1})=e^{A \Delta } \cdot X(t_{i}) + \xi_i,
\end{equation} 
where $\xi_i$ are $d$-dimensional Gaussian vectors with null mean and variance $C(\Delta)$, where the matrix $C(t)$ follows the dynamics of the matrix-valued ODE
\begin{equation}\label{17b}
	\dot{C}(t)=AC(t)+C(t)A^{'} + BB^{'},
\end{equation}
see \cite{9}. Moreover, since the non-linear term $G$  depends only on the component $\textbf{Q}$, the exact path of Subsystem \eqref{Sub2} is obtained through
\begin{equation}\label{Exact_ODE}
	X(t_{i+1})=X(t_i) + \left( \begin{array}{c} 0_d\\\Delta  G(Q(t_i);\theta)  \end{array} \right).
\end{equation}
We apply the Strang approach given by
\begin{equation} \label{Strang}
	(\varphi^b_{\Delta/2} \circ \varphi^a_{\Delta} \circ \varphi^b_{\Delta/2})(x), \quad x \in \mathbb{R}^n,
\end{equation}
where $\varphi_t^a$ and $\varphi_t^b$ denote the exact solutions \eqref{Exact_SDE} and \eqref{Exact_ODE} of \eqref{Sub1}  and \eqref{Sub2}, respectively. Hence, given $X(t_i)$, we obtain the next value $X(t_{i+1})$ by applying the following three steps:
\begin{algorithmic}[1]	
	\State $X_b=X(t_i)+ \left( \begin{array}{c} 0_d\\ \frac{\Delta}{2}  G(Q(t_i);\theta)  \end{array} \right)$
	\State $X_a=e^{A \Delta } \cdot X_b + \xi_i$
	\State $X(t_{i+1})=X_a+ \left( \begin{array}{c} 0_d\\ \frac{\Delta}{2}  G(Q_a;\theta)  \end{array} \right)$
\end{algorithmic} 	

The derivation of the two subsystems is not unique. For example, another possibility is to combine the stochastic term with the non-linear part, yielding the subsystems
\begin{equation} \label{Sub3}
	d\begin{pmatrix} Q(t)\\ P(t) \end{pmatrix}=
	\underbrace{\begin{pmatrix} \nabla_P H(Q(t),P(t)) \\ -\nabla_Q H(Q(t),P(t)) - 2 \Gamma_\theta P(t) \end{pmatrix}}_{f^{[1]}(X(t);\theta)}dt,
\end{equation} 
\begin{equation} \label{Sub4}
	d\begin{pmatrix} Q(t) \\ P(t)  \end{pmatrix}=\underbrace{\begin{pmatrix} 0_d \\ G(Q(t);\theta) \end{pmatrix}}_{f^{[2]}(Q(t);\theta)}dt+
	\underbrace{\left( \begin{array}{c}\mathbb{O}_d\\\Sigma_\theta \end{array} \right)}_{\mathcal{G}^{[2]}(\theta)}dW(t).
\end{equation}
The exact path of \eqref{Sub3} is given by
\begin{equation}\label{Exact_SDEb}
	X(t_{i+1})=e^{A \Delta } \cdot X(t_{i}),
\end{equation} 
while the exact path of \eqref{Sub4} is obtained through
\begin{equation}\label{Exact_ODEb}
	X(t_{i+1})=\left( \begin{array}{c} Q({t_{i}})\\P({t_{i}})+\Delta  G(Q(t_i);\theta) + \Sigma_\theta \cdot \xi_i \end{array} \right),
\end{equation}
where $\xi_i$ are $d$-dimensional Gaussian vectors with null mean and variance $\Delta\mathbb{I}_d$. The Strang approach is now given by
\begin{equation} \label{Strang_2}
	(\varphi^c_{\Delta/2} \circ \varphi^d_{\Delta} \circ \varphi^c_{\Delta/2})(x), \quad x \in \mathbb{R}^n,
\end{equation}
where $\varphi_t^c$ and $\varphi_t^d$ denote the exact solutions \eqref{Exact_SDEb} and \eqref{Exact_ODEb} of \eqref{Sub3} and \eqref{Sub4}, respectively. Thus, given $X(t_i)$, the next value $X(t_{i+1})$ is obtained via:
\begin{algorithmic}[1]	
	\State $X_c=e^{A \frac{\Delta}{2} } \cdot X(t_i)$
	\State $X_d=X_c+ \left( \begin{array}{c} 0_d\\ \Delta  G(Q_c;\theta) + \Sigma_\theta \cdot \xi_i \end{array} \right)$
	\State $X(t_{i+1})=e^{A \frac{\Delta}{2} } \cdot X_d$
\end{algorithmic} 	

\subsection{Implementation details}
\label{subsec:3:3}
The ABC procedure is coded  in the computing environment \textbf{R} \citep{R}, using the package \textbf{Rcpp} \citep{Rcpp}, which offers a seamless integration of \textbf{R} and \emph{C++}, drastically reducing the computational time of the algorithms. 
The code is then parallelised using the R-packages \texttt{foreach} and \texttt{doParallel}, taking advantage of the {\em for-loop} in the algorithm. All simulations are run on the HPC cluster \texttt{RADON1}, a high-performing multiple core cluster located at the Johannes Kepler University Linz. 
To obtain smoothed periodogram estimates, we apply the R-function \texttt{spectrum}. It requires the specification of a smoothing parameter \texttt{span}. In all our experiments, we use \texttt{span}~ $=5T$. In addition, we avoid using a logarithmic scale by setting the \texttt{log} parameter to \lq\lq no\rq\rq. To obtain kernel estimates of the invariant density, we apply the R-function \texttt{density}. Here, we use the default value for the smoothing bandwidth \texttt{bw} and set the number of points at which the invariant density has to be estimated to \texttt{n}$=10^3$. The invariant spectral density is estimated at the default values of the \texttt{spectrum} function. A sample code is publicly available on github 
at \url{https://github.com/massimilianotamborrino/sdbmpABC}.

\section{Validation of the proposed ABC method when exact simulation is possible}
\label{sec:4}
In this section, we illustrate the performance of the proposed ABC approach on a model problem (weakly damped stochastic harmonic oscillator) of Hamiltonian type \eqref{Hamiltonian} with vanishing non-linear displacement term $G \equiv 0$. Linear SDEs of this type reduce to \eqref{narrow_SDE} and allow for an exact simulation of sample paths through \eqref{Exact_SDE}. Therefore, we can apply the Spectral Density-Based ABC Algorithm \ref{Algorithm_Standard} (i) under the optimal condition of exact, and thus $\eta_{\textbf{Y}_\theta}$-preserving data generation. Its performance is illustrated in Subsection \ref{subsec:4:2}. To investigate how the numerical error in the synthetic data generation impinges on the ABC performance, in Subsection \ref{subsec:4:3} we compare $\pi_{\textrm{ABC}}(\theta|y)$ with the posterior densities $\pi_{\textrm{ABC}}^{\text{num}}(\theta|y)$ and $\pi_{\textrm{ABC}}^{\text{e}}(\theta|y)$ obtained from Algorithm \ref{Algorithm_Standard} (ii) and (iii) using the measure-preserving numerical splitting scheme \eqref{Strang_2} and  the non-preserving Euler-Maryuama method \eqref{EM}, respectively.

\subsection{Weakly damped stochastic harmonic oscillator: The model and its properties}
\label{subsec:4:1}
We investigate the $2$-dimensional Hamiltonian type SDE
\begin{equation}
	\begin{footnotesize}
		d \begin{pmatrix}
			Q(t) \\
			P(t) 
		\end{pmatrix}
		=
		\begin{pmatrix}
			P(t) \\
			-\lambda^2Q(t) -2\gamma P(t) 
		\end{pmatrix} dt \ + \
		\begin{pmatrix}
			0 \\
			\sigma
		\end{pmatrix} dW(t),
	\end{footnotesize}
	\label{MP}
\end{equation}
with strictly positive parameters $\gamma$, $\lambda$ and $\sigma$. Depending on the choice of $\gamma$ and $\lambda$, \eqref{MP} models different types of harmonic oscillators, which are common in nature and of great interest in classical mechanics. Here, we focus on the weakly damped harmonic oscillator, satisfying the condition $\lambda^2-\gamma^2>0$. Our goal is to estimate $\theta=(\lambda,\gamma,\sigma)$ assuming that the solution process $\textbf{X}=(\textbf{Q},\textbf{P})^{'}$ is partially observed through the first coordinate, i.e., $\textbf{Y}_\theta=\textbf{Q}$. An illustration of the performance of Algorithm \ref{Algorithm_Standard} (i) for the critically damped case satisfying $\lambda^2-\gamma^2=0$, when only the second coordinate is observed, is reported in the supplementary material.
The solution process $\textbf{X}$ of SDE \eqref{MP} is normally distributed according to
\begin{equation*}
	X(t) \sim \eta_\textbf{X}(t):= \mathcal{N}\Big(e^{At} \cdot \mathbb{E}[X_0], \ \textrm{Var}[e^{At} \cdot X_0] + C(t)\Big),
\end{equation*}
with $A$ and $C$ introduced in \eqref{narrow_SDE} and \eqref{17b}, respectively.
The invariant distribution $\eta_\textbf{X}$ of the solution process $\textbf{X}$ is given by
\begin{equation*}
	\eta_\textbf{X}=\lim\limits_{t \to \infty} \eta_\textbf{X}(t)= 
	\mathcal{N}\left(\begin{pmatrix}
		0  \\
		0  
	\end{pmatrix},\begin{pmatrix}
		\frac{\sigma^2}{4\gamma \lambda^2} & 0 \\
		0 & \frac{\sigma^2}{4\gamma} 
	\end{pmatrix}\right).
\end{equation*}
Consequently, the structural property $\eta_{\textbf{Y}_\theta}$ of the output process $\textbf{Y}_\theta$ becomes
\begin{equation} \label{invD}
	\eta_{\textbf{Y}_\theta} = \mathcal{N}\left(0,\frac{\sigma^2}{4\gamma\lambda^2}\right),
\end{equation}
and the stationary dependency is captured by the autocovariance function
\begin{equation*}
	r_\theta(\Delta) =
	\frac{\sigma^2}{4\lambda^2}e^{-\gamma \Delta} \left[\frac{1}{\gamma}\cos(\kappa\Delta)+\frac{1}{\kappa}\sin(\kappa\Delta)\right],
\end{equation*}
where $\kappa=\sqrt{\lambda^2-\gamma^2}$.

\subsection{Validation of the Spectral Density-Based ABC Algorithm \ref{Algorithm_Standard} (i)}
\label{subsec:4:2}
To compare the performances of Algorithm \ref{Algorithm_Standard} (i)-(iii)  on the same data, we consider the same $M=10$ observed paths simulated with the exact scheme \eqref{Exact_SDE}, using a time step $\Delta =10^{-2}$ over a time interval of length $T=10^3$. As true parameters for the simulation of the reference data, we choose 
\begin{equation*}
	\theta^t=(\lambda^t, \gamma^t,\sigma^t)=(20,1,2).
\end{equation*}
We use the exact simulation scheme \eqref{Exact_SDE} to generate $N=2\cdot 10^6$ synthetic datasets {in $[0,T]$} and with the same time step as the observed data. We choose independent uniform priors, in particular, 
\begin{equation*}
	\lambda\sim U(18,22), \ \gamma\sim U(0.01,2.01), \ \sigma\sim U(1,3).
\end{equation*}
The tolerance level $\epsilon$ is chosen as the $0.05^{\text{th}}$ percentile of the calculated distances. Hence, we keep $10^3$ of all the sampled values for $\theta$. In all the considered examples (see also the supplementary material), the performance of the ABC algorithms for the estimation of the parameters of SDE \eqref{MP} does not improve when incorporating the information of the invariant densities into the distance \eqref{weight}. This is because the mean of the invariant distribution \eqref{invD} is zero. Hence, to reduce the computational cost, we set $w=0$ and base our distance only on the invariant spectral density, estimated by the periodogram.

\begin{figure}[H]
	\centering	
	\subfigure{\includegraphics[width=1.0\textwidth]{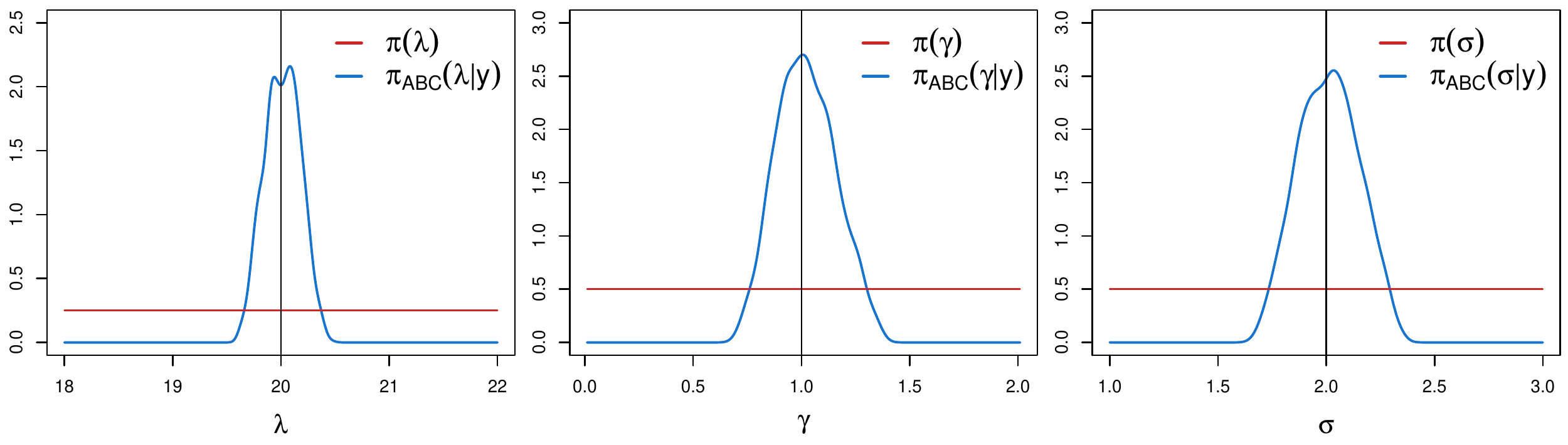}}	\subfigure{\includegraphics[width=1.0\textwidth]{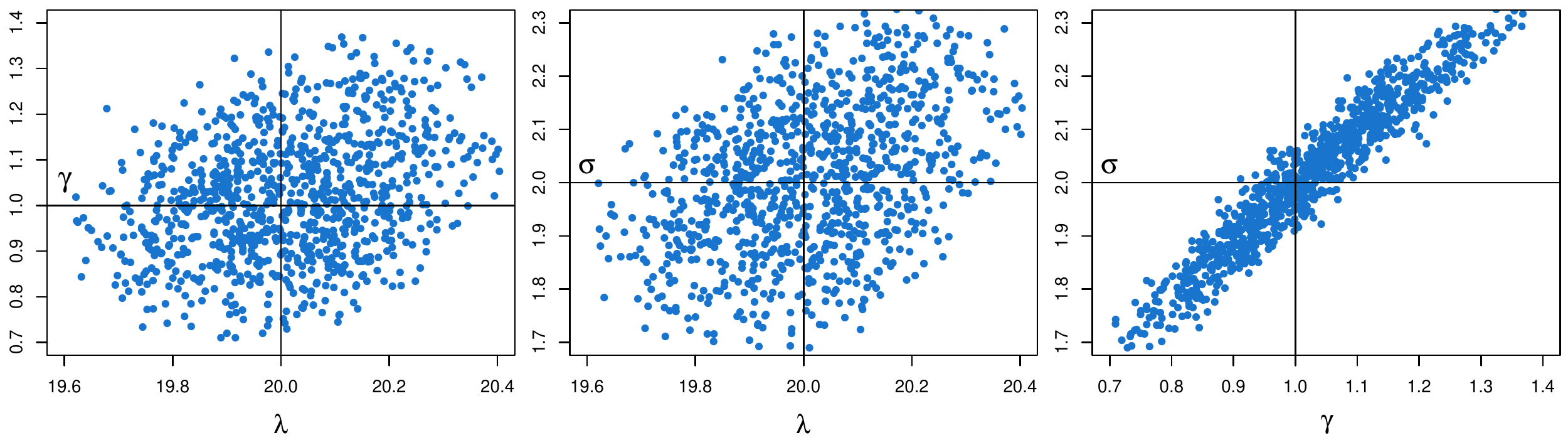}}
	\caption{Top panels: ABC marginal posterior densities $\pi_{\textrm{ABC}}(\theta_j|y)$ (blue lines) of $\theta=(\lambda,\gamma,\sigma)$ of the weakly damped stochastic harmonic oscillator \eqref{MP} and uniform priors (red lines). The posteriors are obtained from Algorithm \ref{Algorithm_Standard} (i). The vertical lines represent the true parameter values. Lower panels: Pairwise scatterplots of the kept ABC posterior samples}
	\label{3_Par_MP3}
\end{figure}

Figure \ref{3_Par_MP3} (top panels) shows the marginal ABC posterior densities $\pi_{\textrm{ABC}}(\theta_j|y)$ (blue lines) and their flat uniform priors $\pi(\theta_j)$ (red lines). The proposed ABC Algorithm \ref{Algorithm_Standard} (i) provides marginal posterior densities centred around the true values $\theta^t$, represented by the black vertical lines. The posterior means 
are given by
\begin{equation*}
	(\hat{\lambda}_{\textrm{ABC}},\hat{\gamma}_{\textrm{ABC}},\hat{\sigma}_{\textrm{ABC}})=(20.015,1.022,2.011).
\end{equation*}
In the lower panels of Figure \ref{3_Par_MP3}, we report the pairwise scatterplots of the kept ABC posterior samples. Note that, since the kept values of $\lambda$ are uncorrelated with those of the other parameters, the support of the obtained marginal posterior density 
is approximately the same as when estimating only $\theta=\lambda$ or $\theta=(\lambda,\gamma)$ (cf. supplementary material). Vice versa, since the kept ABC posterior samples of the parameters $\gamma$ and $\sigma$ are correlated, the support of $\pi_{\textrm{ABC}}(\gamma|y)$ is larger than that obtained when estimating $\theta=(\lambda,\gamma)$. Despite this correlation, Algorithm \ref{Algorithm_Standard} (i) allows for a successful inference of all the three parameters. 

\subsection{Validation of the Spectral Density-Based and Measure-Preserving ABC Algorithm \ref{Algorithm_Standard} (ii)}
\label{subsec:4:3}
\begin{figure}[H]
	\centering	
	\subfigure{\includegraphics[width=1.0\textwidth]{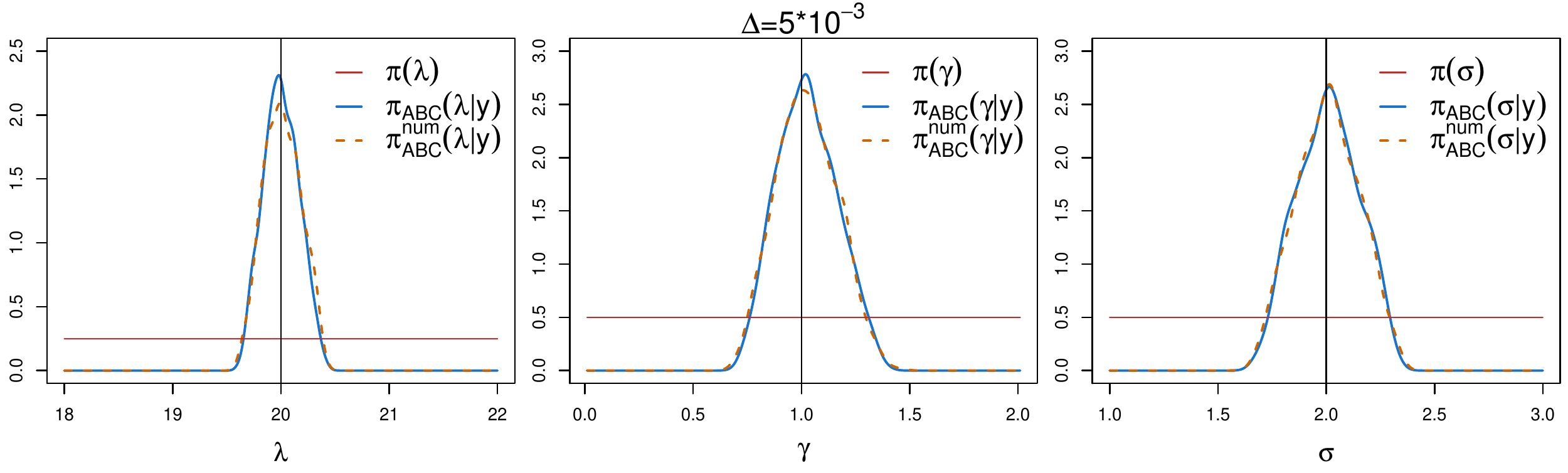}}	
	\subfigure{\includegraphics[width=1.0\textwidth]{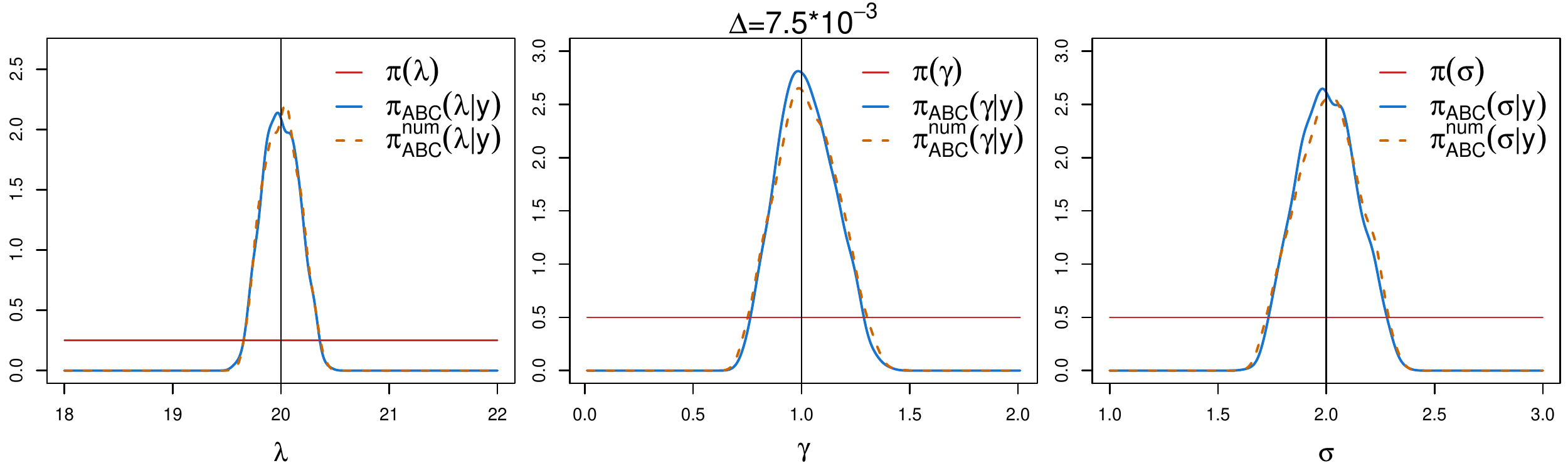}}
	\subfigure{\includegraphics[width=1.0\textwidth]{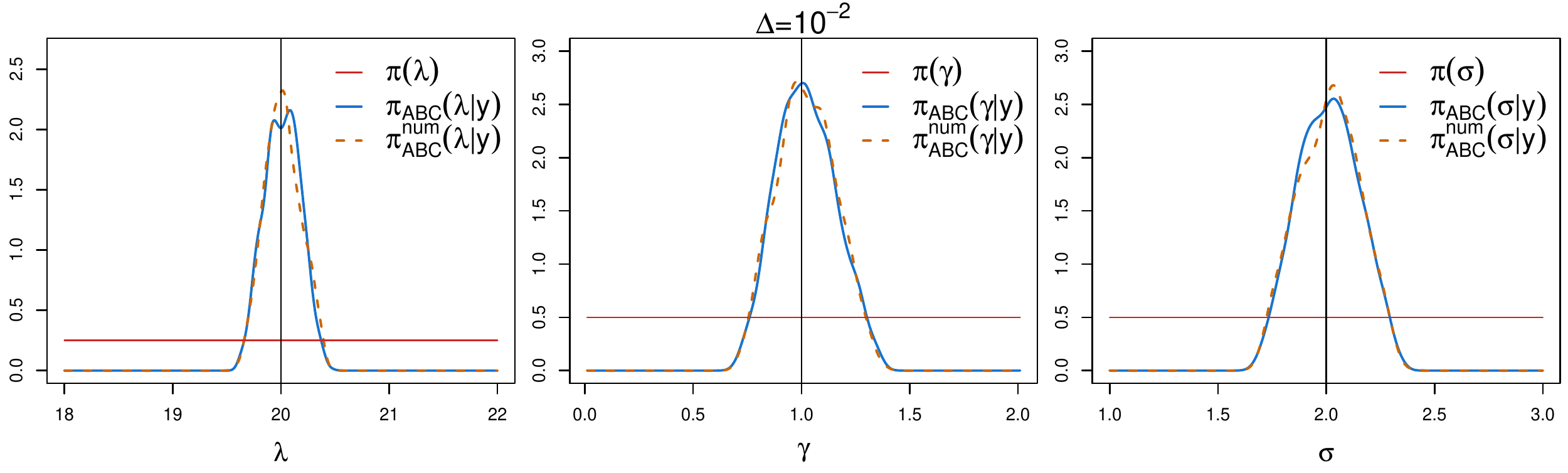}}	
	\caption{ABC marginal posterior densities of $\theta=(\lambda,\gamma,\sigma)$ of the weakly damped stochastic harmonic oscillator \eqref{MP} obtained from Algorithm \ref{Algorithm_Standard} (i) with the exact simulation method \eqref{Exact_SDE} (blue solid lines) and Algorithm \ref{Algorithm_Standard} (ii) combined with the splitting scheme \eqref{Strang_2} (orange dashed lines) for different choices of the time step $\Delta$. In particular, $\Delta= 5\cdot 10^{-3}$ (top panels), $7.5 \cdot 10^{-3}$ (middle panels) and $10^{-2}$ (lower panels). The red horizontal lines denote the uniform priors and the black vertical lines the true parameter values}
	\label{ABC2ABC3}
\end{figure}

In Figure \ref{ABC2ABC3}, we report the approximated marginal posteriors $\pi_{\textrm{ABC}}(\theta_j|y)$ (blue solid lines) and $\pi_{\textrm{ABC}}^\textrm{num}(\theta_j|y)$ (orange dashed lines) obtained with the same priors, $\epsilon$, $T$, $w$, $M$ and $N$ as before, for different values of the time step $\Delta$. In particular, we choose  $\Delta= 5\cdot 10^{-3}$ (top panels), $\Delta=7.5 \cdot 10^{-3}$ (middle panels) and $\Delta=10^{-2}$ (lower panels). The posteriors obtained from Algorithm \ref{Algorithm_Standard} (ii) successfully targets $\pi_{\textrm{ABC}}(\theta|y)$, even for a time step as large as $\Delta=10^{-2}$. On the contrary, Algorithm \ref{Algorithm_Standard} (iii) is not even applicable. Indeed, the numerical scheme computationally pushes the amplitude of the oscillator towards infinity, resulting in a computer overflow, i.e., $\widetilde{Y}_\theta(t_i) \approx \infty$. Thus, neither $\hat{f}_{\tilde y_\theta}$ nor $\hat{S}_{\tilde y_\theta}$ can be computed and the density $\pi_{\textrm{ABC}}^e(\theta|y)$ cannot be derived.

\begin{figure}[H]
	\includegraphics[width=1.0\textwidth]{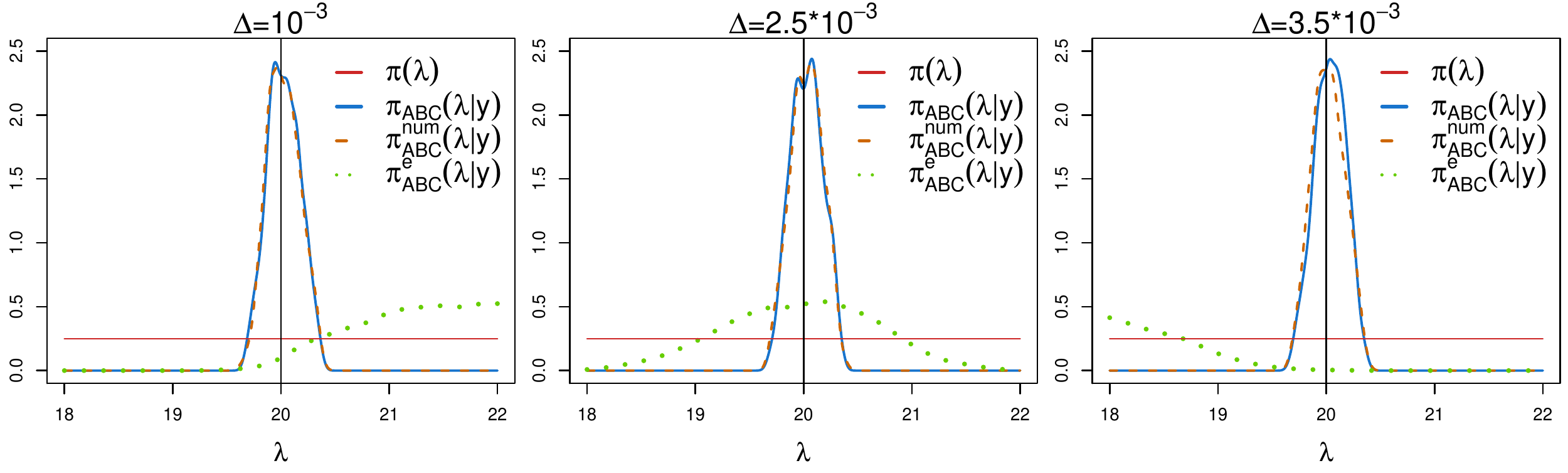}\\
	\caption{ABC posterior densities of $\theta=\lambda$ of the weakly damped stochastic oscillator \eqref{MP} obtained from Algorithm \ref{Algorithm_Standard} (i) using the exact simulation scheme \eqref{Exact_SDE} (blue solid lines), (ii) using the splitting scheme \eqref{Strang_2} (orange dashed lines) and 
		(iii) using the Euler-Maruyama method \eqref{EM} (green dotted lines) for different choices of the time step $\Delta$. The horizontal red lines and the vertical black lines represent the uniform priors and the true parameter values, respectively}
	\label{Plot31_C5}
\end{figure}

As a further illustration of the poor performance of the Euler-Maruyama scheme, even for smaller choices of $\Delta$, we now consider the simplest possible scenario where we only estimate  one parameter, namely $\theta=\lambda$. We set $N=10^5$, $M=10$, $\epsilon=1^{\text{st}}$ percentile and we choose a uniform prior
$\lambda\sim U(10,30)$.
To be able to derive $\pi_{\textrm{ABC}}^{e}(\lambda|y)$, we simulate the synthetic data using the Euler-Maruyama method with the time steps $\Delta=10^{-3}$, $2.5 \cdot 10^{-3}$ and $3.5\cdot 10^{-3}$. Figure \ref{Plot31_C5} shows the three ABC posterior densities $\pi_{\textrm{ABC}}(\theta|y)$ (blue solid lines), $\pi_{\textrm{ABC}}^{\textrm{num}}(\theta|y)$ (orange dashed lines) and $\pi_{\textrm{ABC}}^{e}(\theta|y)$  (green dotted lines) for the different choices of $\Delta$. The horizontal red lines and the black vertical lines denote the uniform prior and the true parameter value, respectively. In all cases, Algorithm  \ref{Algorithm_Standard} (iii) does not lead to a successful inference. In addition, these results are not stable for the different choices of $\Delta$, and the derived ABC posterior density may not even cover the true parameter value. 

\section{Validation of the Spectral Density-Based and Measure-Preserving ABC Algorithm \ref{Algorithm_Standard} (ii) on simulated and real data}
\label{sec:5}
We now illustrate the performance of Algorithm \ref{Algorithm_Standard} (ii) by applying it to 
the stochastic JR-NMM. We rely on the efficient numerical splitting scheme \eqref{Strang} to guarantee measure-preserving synthetic data generation within the ABC framework.
After estimating the parameters from simulated data, we infer them from real EEG data. In the available supplementary material, we illustrate the performance of Algorithm \ref{Algorithm_Standard} (ii) also on the non-linear damped stochastic oscillator, an extended version of the weakly damped harmonic oscillator discussed in Section \ref{sec:4}.

\subsection{The stochastic Jansen and Rit neural mass model}
\label{subsec:5:1}
The stochastic JR-NMM describes the electrical activity of an entire population of neurons through their average properties by modelling the interaction of the main pyramidal cells with the surrounding excitatory and inhibitory interneurons. The model has been reported to successfully reproduce EEG data, and is applied in the research of neurological disorders such as epilepsy or schizophrenia \citep{60,59}. The model is a $6$-dimensional SDE of the form 
\begin{equation}\label{JR-NMM}
d \begin{pmatrix}
Q(t) \\
P(t) 
\end{pmatrix}
=\begin{pmatrix}
P(t) \\
-\Gamma^2Q(t) -2\Gamma P(t) + G(Q(t);\theta)
\end{pmatrix} dt+	\begin{pmatrix}
0 \\
\Sigma_\theta
\end{pmatrix} dW(t),
\end{equation}
where the $6$-dimensional solution process is given by $\textbf{X}=(\textbf{Q},\textbf{P})^{'}$ with the two components 
$\mathbf{Q}=(\mathbf{X_1},\mathbf{X_2},\mathbf{X_3})^{'}$ and 
$\mathbf{P}=(\mathbf{X_4},\mathbf{X_5},\mathbf{X_6})^{'}$.  None of the coordinates of $\textbf{X}$ is directly observed. Only the difference between the second and third coordinates can be measured with EEG-recording techniques, yielding the output process
\begin{equation*}
	\textbf{Y}_\theta=\mathbf{X_2}-\mathbf{X_3}.
\end{equation*}

In \eqref{JR-NMM}, the diagonal diffusion matrix is given by $\Sigma_\theta$=diag$[\sigma_4, \sigma_5, \sigma_6] \in \mathbb{R}^{3\times 3}$ with coefficients $\sigma_i>0$, $i=4,5,6$.
The matrix $\Gamma$=diag$[a,a,b] \in \mathbb{R}^{3\times 3}$ is also diagonal with coefficients $a,b>0$, representing the time constants of the excitatory and inhibitory postsynaptic potentials, respectively. The non-linear displacement term is given by
\begin{equation*}
	G(\mathbf{Q};\theta)=
	\begin{pmatrix} Aa[$Sigm$(\mathbf{X_2}-\mathbf{X_3})] \\ Aa[\mu+C_2$Sigm$(C_1\mathbf{X_1})] \\ Bb[C_4$Sigm$(C_3\mathbf{X_1})]  \end{pmatrix},
\end{equation*}
where the sigmoid function Sigm: $\mathbb{R} \to [0, v_{max}]$ is defined as
\begin{equation*}
	\text{Sigm}(x):=\frac{v_{max}}{1+\text{exp}[r(v_0-x)]},
\end{equation*}
with $v_{max}>0$ referring to the maximum firing rate of the neural populations, $v_0 \in \mathbb{R}$ describing the value for which $50 \ \%$ of the maximum firing rate is attained and $r>0$ denoting the slope of the sigmoid function at $v_0$. The parameters entering in $G$ are $\mu$, $A$, $B$ and $C_i$, $i=1,2,3,4$ $\in \mathbb{R}^+$. The coefficients $A$ and $B$ describe the average excitatory and inhibitory synaptic gain, respectively. The parameters $C_i$ are internal connectivity constants, which reduce to only one parameter $C$, by using the relations $C_1=C$, $C_2=0.8C$, $C_3=0.25C$ and $C_4=0.25C$; see \cite{1}. 

\subsection{Parameter inference from simulated data}
\label{subsec:5:2}
Not all model parameters of the JR-NMM are of biological interest or can be simultaneously identified. For example, the noise coefficients $\sigma_4$ and $\sigma_6$ were introduced mainly for mathematical convenience  in \cite{2}. To guarantee the existence of a unique invariant measure $\eta_{\textbf{X}}$ on $(\mathbb{R}^6,\mathcal{B}(\mathbb{R}^6))$, they are required to be strictly positive. However, from a modelling point of view, only the parameter $\sigma:=\sigma_5$ plays a role. Hence, we fix $\sigma_4=0.01$ and $\sigma_6=1$. The coefficients $A$, $B$, $a$, $b$, $v_0$, $v_{max}$ and $r$ {have been experimentally recorded}; see, e.g., \cite{Jansen1993,1,vanRotterdam1982}. Thus, we fix them according to these values reported, for example, in Table $1$ of \cite{2}. In contrast, the connectivity parameter $C$, which represents the average number of synapses between the neural subpopulations and controls to what extent the main population interacts with the interneurons, {varies} under different physiological constraints. Changing $C$ allows, for example, a transition from $\alpha$-rhythmic activity to epileptic spiking behaviour; see, e.g., \cite{2}. Here, we focus on the $\alpha$-rhythmic activity.  
Since the parameters $\sigma$ and $\mu$ are new in the SDE version \eqref{JR-NMM}, they have not yet been estimated. They can be interpreted as stochastic and deterministic external inputs coming from neighbouring or more distant cortical columns, respectively. Thus, together with the internal connectivity parameter $C$, they are of specific interest. Before inferring $\theta=(\sigma,\mu,C)$, we take into account the coefficients $A$ and $B$ to discuss a model-specific issue of identifiability.

\subsubsection{Identifiability issues: The detection of an invariant manifold, i.e., a set of parameters yielding the same type of data}
\label{subsubsec:5:2:1}
For the original JR-NMM, it has been shown that different combinations of the parameters $A$, $B$ and $C$ yield the same type of output, namely the $\alpha$-rhythmic brain activity. Applying the proposed Spectral Density-Based and Mea\-sure-Preserving ABC Algorithm \ref{Algorithm_Standard} (ii) for the inference of $\theta=(A,B,C)$, with given $\mu=220$ and $\sigma=2000$, we confirm that the same non identifiability arises for the SDE version \eqref{JR-NMM}. We choose $M=30$ observed paths generated assuming
\begin{equation*}
	\theta^t=(A^t,B^t,C^t)=(3.25,22,135),
\end{equation*}
as suggested in the literature \citep{1}.
The reference and synthetic data are generated over a time interval of length $T=200$ and using a time step $\Delta=2 \cdot 10^{-3}$. 
Within the algorithm, we generate $N=2.5 \cdot 10^6$ synthetic datasets. We choose the weight $w$ in \eqref{weight} according to the procedure introduced in Subsection \ref{subsec:2:2} (based on {$L=10^5$} iterations) and fix the tolerance level $\epsilon=0.04^{\text{th}}$ percentile. Further, we choose independent uniform prior distributions, namely 
\begin{equation*}
	A \sim \mathcal{U}(1,10), \ B \sim \mathcal{U}(10,100), \ \ C \sim \mathcal{U}(10,600).
\end{equation*}
\begin{figure}[H]
	\centering	
	\subfigure{\includegraphics[width=1.0\textwidth]{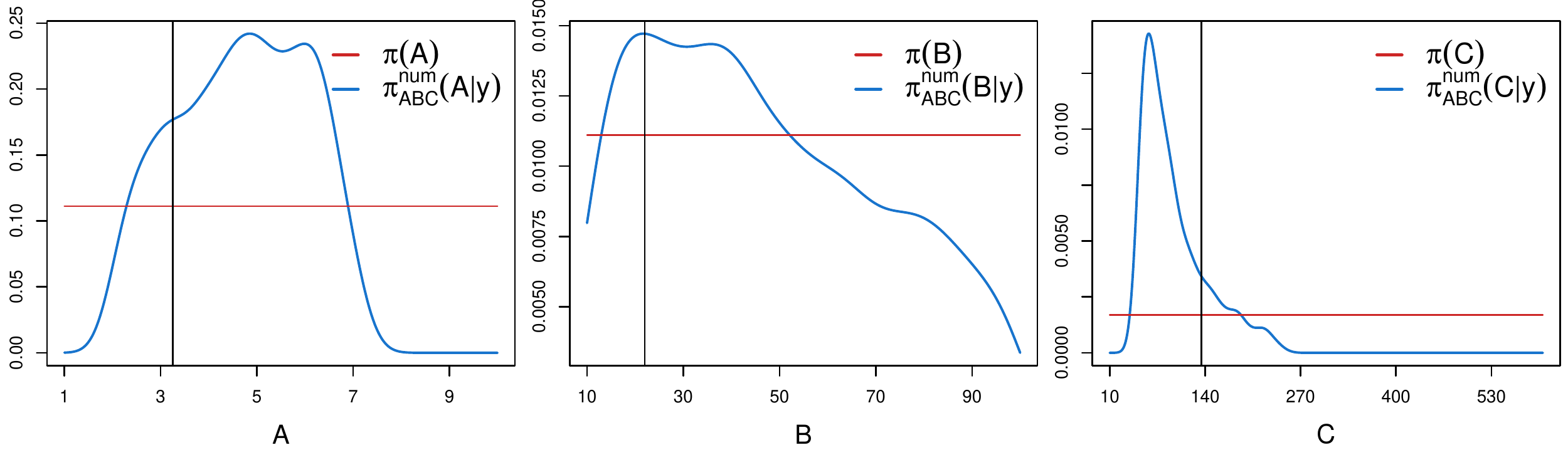}}		
	\subfigure{\includegraphics[width=1.0\textwidth]{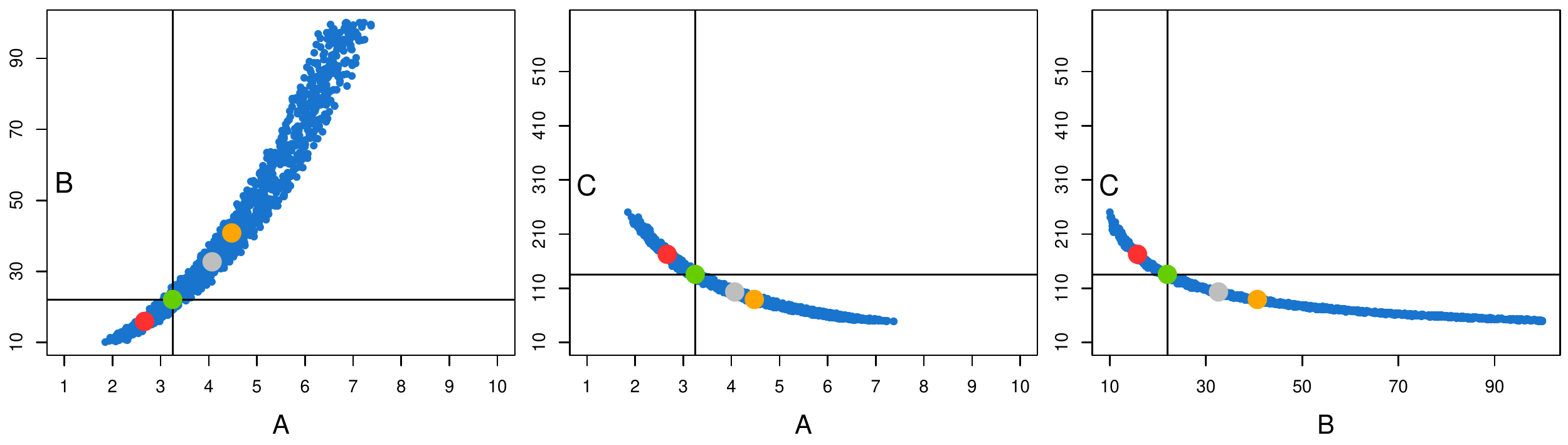}}	
	\subfigure{\includegraphics[width=1.0\textwidth]{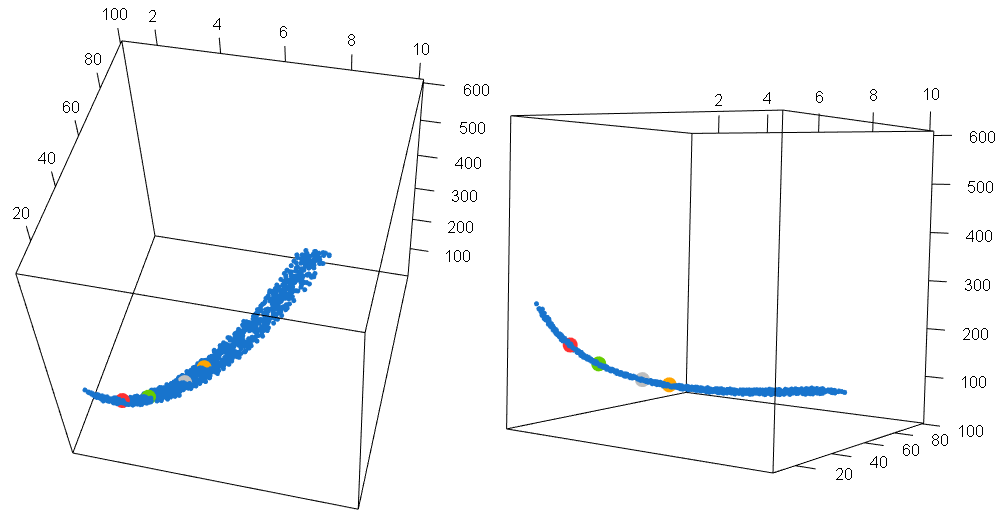}}
	\caption{Top panels: ABC marginal posterior densities $\pi_{\textrm{ABC}}^{\text{num}}(\theta_j|y)$ (blue lines) of $\theta=(A,B,C)$ of the stochastic JR-NMM  \eqref{JR-NMM} obtained from Algorithm \ref{Algorithm_Standard} (ii). The horizontal red lines and the vertical black lines represent the uniform priors and the true parameter values, respectively. Middle panels: Pairwise scatterplots of the kept ABC posterior samples. Lower panels: Two different views of a $3$-dimensional scatterplot of the kept ABC posterior samples within a cuboid formed by the prior. The green dot corresponds to $\theta^{t}$ and the red, orange and grey dots represent highlighted samples from the ABC posterior lying on the invariant manifold} \label{JRNMM_ABC}
\end{figure}

\begin{figure}[H]
	\includegraphics[width=1.0\textwidth]{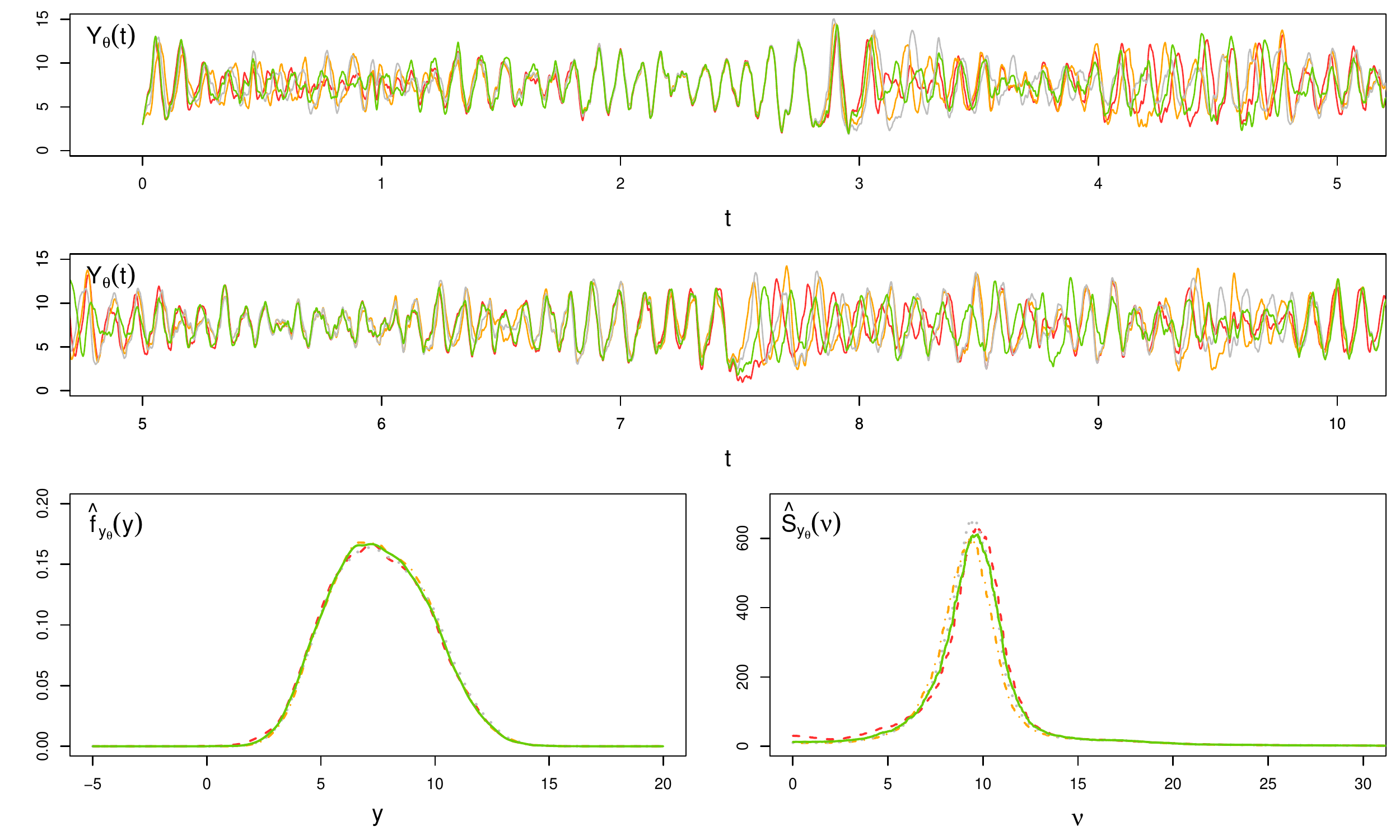}\\
	\caption{Top and middle panel: Four paths of the output process $\textbf{Y}_\theta=\mathbf{X_2}-\mathbf{X_3}$ of the stochastic JR-NMM \eqref{JR-NMM}
		generated under $\theta^t$ (green lines) and with the three highlighted kept ABC posterior samples lying on the invariant manifold of Figure \ref{JRNMM_ABC} (red, orange and grey lines) using the same random numbers. Lower panels: Corresponding estimated invariant densities (left) and estimated spectral densities (right)} \label{manifold}
\end{figure}

Figure \ref{JRNMM_ABC} (top panels) shows the marginal ABC posterior densities $\pi_{\textrm{ABC}}^{\textrm{num}}(\theta_j|y)$ and the uniform prior densities $\pi(\theta_j)$. Clearly, the parameters cannot be inferred simultaneously. The kept ABC posterior values of the parameters $A$, $B$ and $C$ are strongly correlated, as observed in the pairwise scatterplots (middle panels) and in the $3$-dimensional scatterplot (two different views, lower panels). The cuboid covers all possible values for $\theta$ drawn from the prior. After running the ABC algorithm, the kept values of $\theta$ from the ABC posterior form an invariant manifold, in the sense that all the parameter values  $\theta$ lying on this manifold yield similar paths $\tilde{y}_\theta$ of the output process. This is shown  in Figure \ref{manifold}, where we
report four trajectories that have been simulated with the same random numbers but using the parameters $\theta^t$ (green dot in Figure  \ref{JRNMM_ABC}) and three of the kept ABC posterior samples lying on the invariant manifold (red, orange and grey dots in Figure \ref{JRNMM_ABC}). A segment of $T=10$ is split in the top and middle panels. In addition, we visualise the corresponding estimated invariant densities (bottom left) and invariant spectral densities (bottom right). This explains why the parameters $A$, $B$ and $C$ are not simultaneously identifiable from the observed data. 
Since the internal connectivity parameter $C$ has an important neuronal meaning, in the following we assume $A$ and $B$ to be known and infer $\theta=(\sigma,\mu,C)$. The estimation of $\theta=(\sigma,\mu)$ when $C$ is known is reported in the supplementary material.

\subsubsection{Inference of $\theta=(\sigma,\mu,C)$}
\label{subsubsec:5:2:2}
\begin{figure}[H]
	\includegraphics[width=1.0\textwidth]{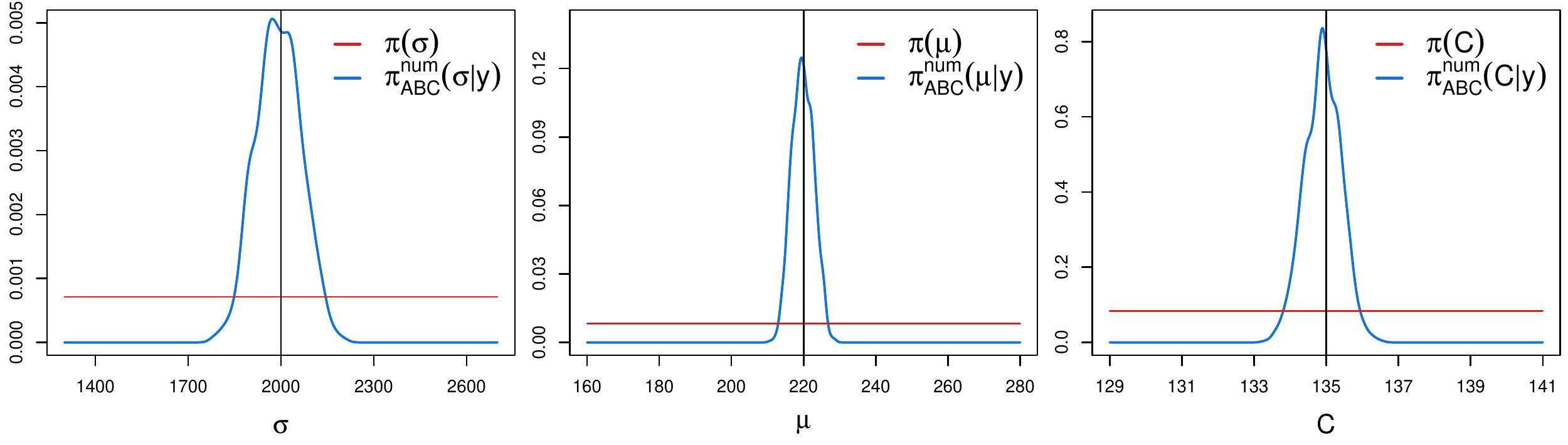}\\
	\caption{ABC marginal posterior densities $\pi_{\textrm{ABC}}^{\text{num}}(\theta_j|y)$ (blue lines) of $\theta=(\sigma,\mu,C)$ of the stochastic JR-NMM \eqref{JR-NMM} obtained from Algorithm \ref{Algorithm_Standard} (ii). The horizontal red lines and the vertical black lines represent the uniform priors and the true  parameter values, respectively} \label{JRNMM_sigmuC}
\end{figure}

Now, we keep the same ABC setting as before and choose independent uniform priors $\pi(\theta_j)$ according to 
\begin{equation*}
	\sigma \sim \mathcal{U}(1300,2700), \
	\mu \sim \mathcal{U}(160,280), \
	C \sim \mathcal{U}(129,141).
\end{equation*}
The reference data are simulated under
\begin{equation*}
	\theta^t=(\sigma^t,\mu^t,C^t)=(2000,220,135).
\end{equation*}
In Figure \ref{JRNMM_sigmuC}, we report the marginal ABC posterior densities $\pi_{\textrm{ABC}}^{\textrm{num}}(\theta_j|y)$ (blue lines), the uniform prior densities $\pi(\theta_j)$ (red lines) and the true parameter values $\theta^t$ (black vertical lines). We obtain unimodal posterior densities, centred around the true parameter values. The posterior density of $\sigma$ is slightly broader compared to that obtained when $C$ is known (cf. Figure 19 of the supplementary material). This results from a weak correlation that we detect  among the kept ABC posterior samples of the parameters $\sigma$ and $C$ (figures not reported). The posterior means are equal to
\begin{equation*}
	(\hat\sigma_{\textrm{ABC}}^{\textrm{num}},\hat\mu_{ABC}^{\textrm{num}},\hat{C}_{\textrm{ABC}}^{\textrm{num}})=(1992.371,219.792,134.904),
\end{equation*}
and are thus close to $\theta^t$. These results suggest an excellent performance of the proposed Spectral Density-Based and Measure-Preserving ABC Algorithm \ref{Algorithm_Standard} (ii). 

Similar satisfactory results are obtained even when adding a fourth parameter, for example, when inferring $\theta=(\sigma,\mu,C,b)$ (cf. Figure $20$ of the supplementary material). When applying Algorithm \ref{Algorithm_Standard} (ii) to real EEG data (cf. Figure $21$ of the supplementary material), the marginal posterior for $b$ is centred around the value $b=50$, which is that reported in the literature. Due to the existence of underlying invariant manifolds, identifiability issues, similar to those reported in Figure \ref{JRNMM_ABC}, arise when adding further or other coefficients, revealing model-specific issues for the stochastic JR-NMM. 

\begin{figure}[H]
	\centering	
	\subfigure{\includegraphics[width=1.0\textwidth]{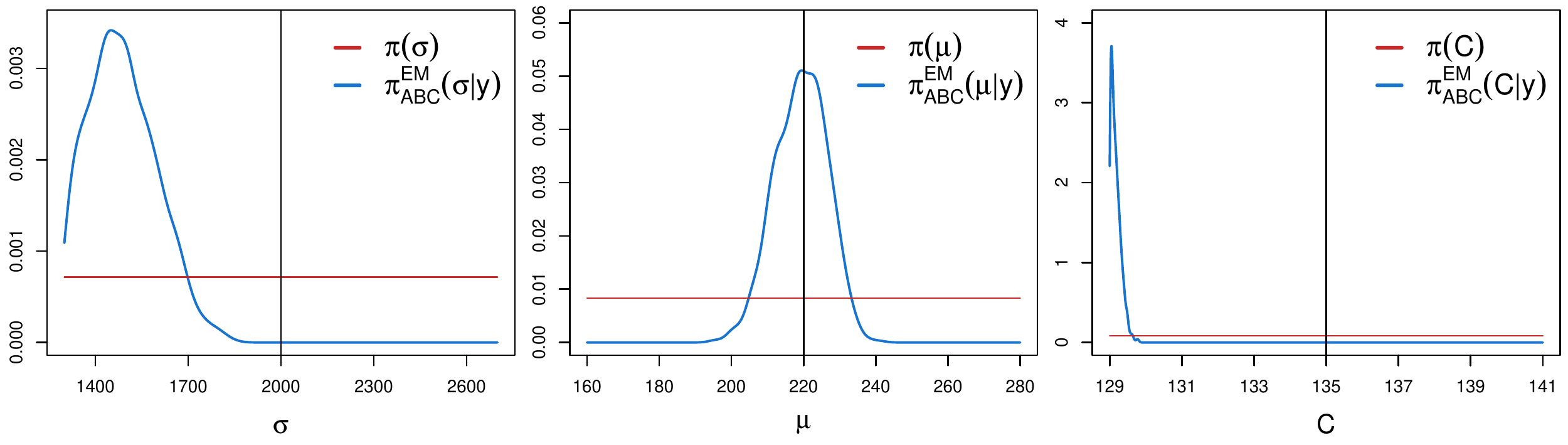}}			
	\subfigure{\includegraphics[width=0.5\textwidth]{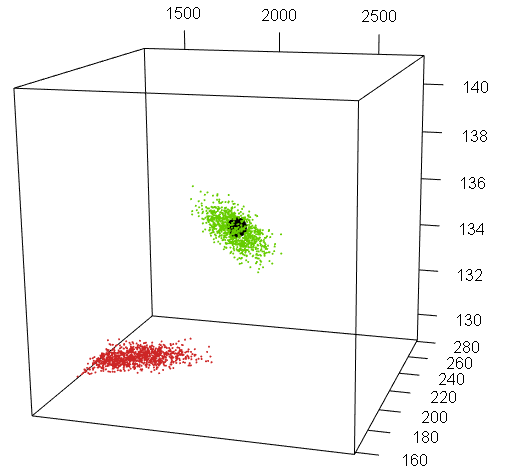}}	
	\caption{Top panels: Marginal ABC posterior densities $\pi_{\textrm{ABC}}^e(\theta_j|y)$  (blue lines) of $\theta=(\sigma,\mu,C)$ of the stochastic JR-NMM \eqref{JR-NMM} obtained from Algorithm \ref{Algorithm_Standard} (iii) using the non-preservative Euler-Maruyama scheme \eqref{EM}. The horizontal red lines and the vertical black lines represent the uniform priors and the true  parameter values, respectively. Lower panel: 3-dimensional scatterplot of the kept ABC posterior samples using Algorithm \ref{Algorithm_Standard} (ii) (green dots; see the previous results reported in Figure \ref{JRNMM_sigmuC}) and  Algorithm \ref{Algorithm_Standard} (iii) (red dots). The cuboid is formed by the prior. The black dot corresponds to $\theta^{t}$} \label{JRNMM_sigmuC_EM}
\end{figure}

To illustrate again the importance of the structure-preservation within the ABC method, we now apply Algorithm \ref{Algorithm_Standard} (iii) combined with the Euler-Maruyama scheme \eqref{EM}. We use the same conditions as before, except for a smaller time step $\Delta=10^{-4}$ used for the generation of the observed reference data with the Euler-Maruyama method aiming for a realistic data structure. In Figure \ref{JRNMM_sigmuC_EM}, we report the marginal ABC posterior densities $\pi_{\textrm{ABC}}^e(\theta_j|y)$ (top panels) and the uniform prior densities. In the $3$-dimensional scatterplot of Figure \ref{JRNMM_sigmuC_EM} (lower panel), the green dots in the middle of the cuboid represent the kept ABC posterior samples when applying Algorithm \ref{Algorithm_Standard} (ii) (see the previous results reported in Figure \ref{JRNMM_sigmuC}), which are nicely spread-out around the true parameter vector $\theta^t$ (black dot). The red dots correspond to the kept ABC posterior samples from $\pi_{\textrm{ABC}}^e(\theta|y)$. Hence, Algorithm \ref{Algorithm_Standard} (iii) based on the Euler-Maruyama scheme provides a posterior that is far off from the true parameter vector.

\subsection{Parameter inference from real EEG data}
\label{subsec:5:3}
\begin{figure}[H]
	\includegraphics[width=1.0\textwidth]{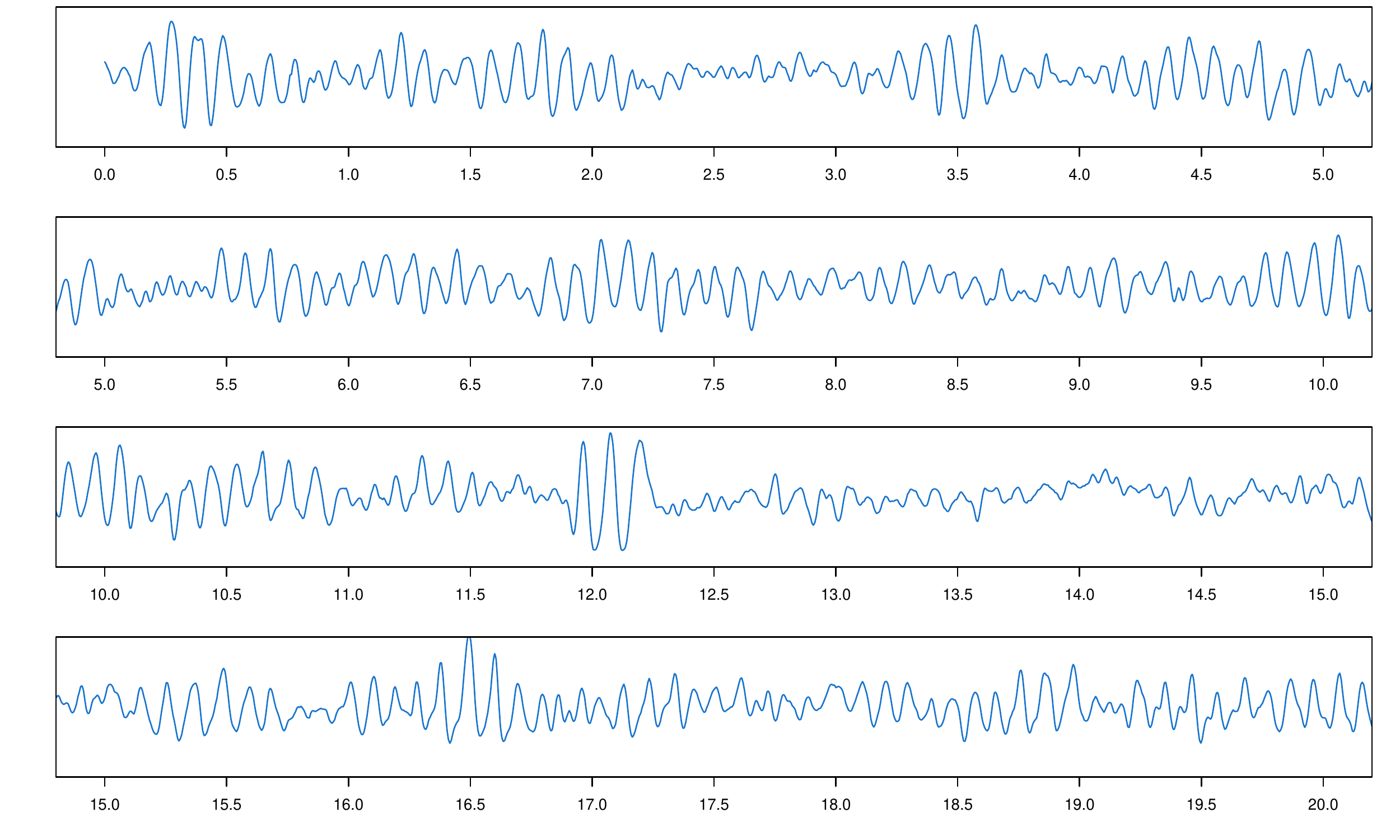}
	\caption{Visualisation of the first $20$ seconds of one of the used $\alpha$-rhythmic EEG segments recorded with a sampling rate of $173.61$ Hz, i.e., $\Delta\approx 5.76$ $\textrm{ms}$} \label{EEG_segment}
\end{figure}

Finally, we use the Spectral Density-Based and Measure-Preserving ABC Algorithm \ref{Algorithm_Standard} (ii) to estimate the parameter vector $\theta=(\sigma,\mu,C)$ of the stochastic JR-NMM from real EEG recordings. We use $M=3$ $\alpha$-rhythmic recordings, rescaled to a realistic range. The EEG data were sampled according to a sampling rate of 173.61 Hz, i.e., a time step $\Delta$ of approximately $5.76\ \textrm{ms}$ over a time interval of length $T=23.6$
$\textrm{s}$. All measurements were carried out with a standardised electrode placement scheme;  see \cite{7} for further information on the data\footnote{The data are  available on: \url{http://ntsa.upf.edu/downloads/andrzejak-rg-et-al-2001-indications-nonlinear-deterministic-and-finite-dimensional}}. Figure \ref{EEG_segment} shows the first $20$ seconds of one of the observed EEG datasets. Here, we  simulate $N=5 \cdot 10^6$ synthetic paths from the output process of the stochastic JR-NMM \eqref{JR-NMM} over the same time interval $T$ as the real data, with a time step $\Delta=2 \cdot 10^{-3}$ and  $\epsilon=0.02^{\text{nd}}$ percentile. We choose independent uniform priors $\pi(\theta_j)$ according to
\begin{equation*}
	\sigma \sim \mathcal{U}(500,3500), \
	\mu \sim \mathcal{U}(70,370), \ 
	C \sim \mathcal{U}(120,150).
\end{equation*}

\begin{figure}[H]
	\includegraphics[width=1.0\textwidth]{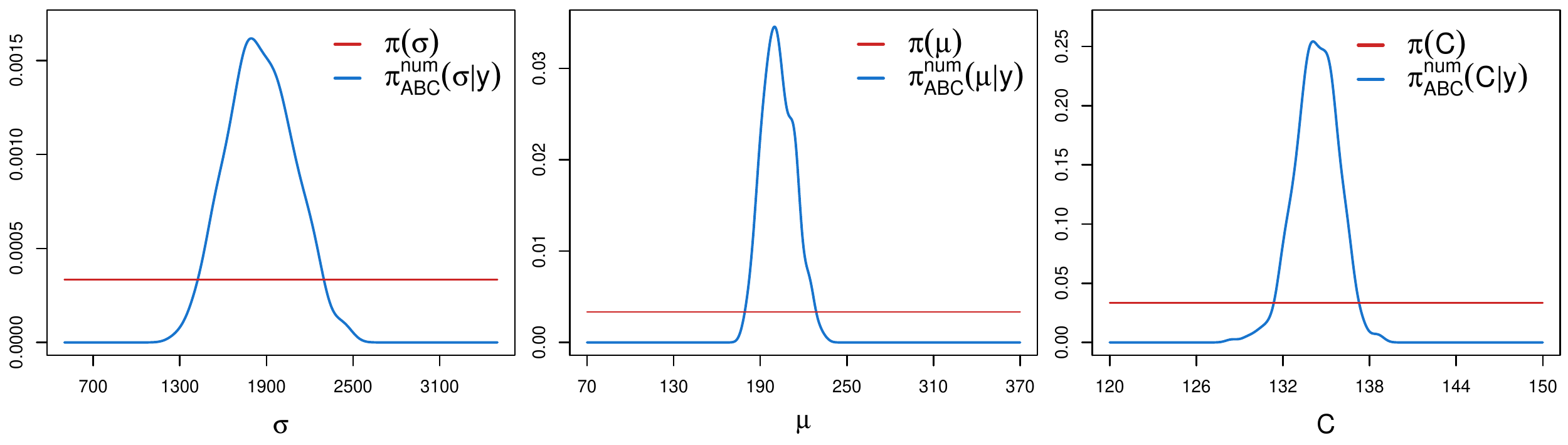}\\
	\caption{Marginal ABC posterior densities $\pi_{\textrm{ABC}}^{\text{num}}(\theta_j|y)$ (blue lines) of $\theta=(\sigma,\mu,C)$ of the stochastic JR-NMM \eqref{JR-NMM} fitted on real EEG data using Algorithm \ref{Algorithm_Standard} (ii). The red lines correspond to the uniform priors} \label{results_EEG} 
\end{figure}

\vspace{-1\baselineskip}
Figure \ref{results_EEG} shows the resulting marginal ABC posterior densities $\pi_{\textrm{ABC}}^\textrm{num}(\theta_j|y)$ and the uniform prior densities $\pi(\theta_j)$. All ABC marginal posteriors are unimodal, with means  given by
\begin{equation*}\label{hat_theta}
	(\hat{\sigma}_{\textrm{ABC}}^\textrm{num},\hat{\mu}_{ABC}^\textrm{num},\hat{C}_{\textrm{ABC}}^\textrm{num})=(1859.211,202.547,134.263).
\end{equation*}
Since $\mu$ and $\sigma$ have not been estimated before, we cannot compare the obtained results with those available in the literature. The ABC posterior density for $C$ is centred around $C=135$ that is the reference literature value for $\alpha$-rhythmic EEG data. 

\begin{figure}[H]
	\includegraphics[width=1.0\textwidth]{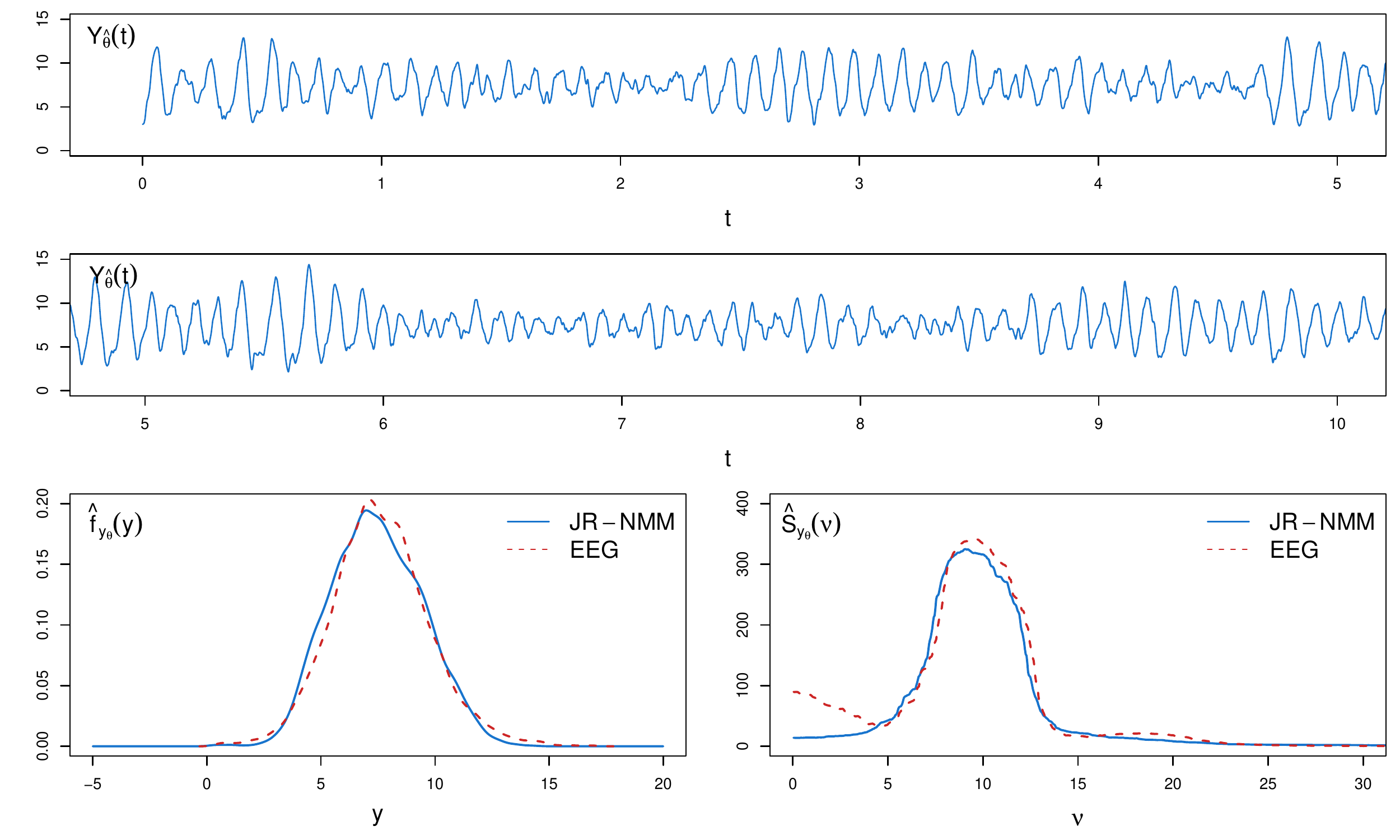}\\
	\caption{Top and middle panel: Visualisation of 10 $s$ of a sample path of the stochastic JR-NMM \eqref{JR-NMM} generated with the numerical splitting scheme \eqref{Strang} using $(\hat\sigma_{\textrm{ABC}}^{\textrm{num}},\hat\mu_{ABC}^{\textrm{num}},\hat{C}_{\textrm{ABC}}^{\textrm{num}})$, the marginal ABC posterior means derived from Algorithm \ref{Algorithm_Standard} (ii). The chosen time step is $\Delta=2 \cdot 10^{-3}$ and $T=23.6$ $s$. Lower panels: Corresponding estimated invariant density (solid blue line, left) and invariant spectral density (solid blue line, right) plotted against those estimated from the real EEG dataset shown in Figure \ref{EEG_segment} (red dashed lines)} 
	\label{Real_vs_Sim} 
\end{figure}

In Figure \ref{Real_vs_Sim}, we report the first $10$ seconds of a trajectory of the output process of the fitted stochastic JR-NMM \eqref{JR-NMM}, generated with the numerical splitting scheme \eqref{Strang}, choosing $\Delta=2 \cdot 10^{-3}$ and $T=23.6$. Note how the path shows a similar oscillatory behaviour as in Figure \ref{EEG_segment}. This is confirmed by noting the satisfactory matches between the invariant densities (bottom left) and the invariant spectral densities (bottom right) estimated from the EEG recording (red dashed lines) and from the fitted model (blue solid lines). The match is poor only for low frequencies of the invariant spectral density, even when choosing broader priors. This may result from a lack of fit of the JR-NMM or of stationarity in the considered EEG data. A deeper investigation of the model and of its ability in reproducing real EEG data is currently under investigation, but it is out of the scope of this work. 

\section{Conclusion}
\label{sec:6}
When performing parameter inference through ABC, crucial and non-trivial tasks are to propose suitable summary statistics and distances to compare the observed and the synthetic datasets. When the underlying models are stochastic, repeated simulations from the same parameter setting yield different outputs, making the {comparison} between the observed and the synthetic data more difficult.   
To derive summary statistics that are less sensitive to the intrinsic randomness of the stochastic model, we propose to map the data to their invariant density and invariant spectral density, estimated by a kernel density estimator and a smoothed periodogram, respectively. By doing this, different trajectories of the output process are mapped to the same objects only when they are generated from the same underlying parameters, provided that all parameters are simultaneously identifiable. These transformations are based on the existence of an underlying invariant measure for the model,  fully characterised by the parameters. A necessary condition of ABC, and of all other simulation-based methods, is the ability to generate data from the model. This is often taken for granted but, in general, it is not the case. Indeed, exact simulation is rarely possible and property-preserving numerical methods have to be derived. 

The combination of the measure-preserving numerical splitting schemes and the use of the spectral density-based distances in the ABC algorithm leads to a  successful inference of the parameters, as illustrated on stochastic Hamiltonian type equations. We validated the proposed ABC approach on both linear model problems, allowing for an exact simulation of the synthetic data, and  non-linear problems, including an application to real EEG data. Our choice of the crucial ingredients (summary statistics and distances based on the underlying invariant distribution and a measure-preserving numerical method) yields excellent results even when applied to ABC in its basic acceptance-rejection form. However, they can be directly applied to more advanced ABC algorithms.
In contrast, the ABC method based on the Euler-Maruayma scheme drastically fails. Its performance may improve for \lq\lq small  enough\rq\rq \ time steps. However, there is a trade-off between the runtime and the acceptance performance of Algorithm \ref{Algorithm_Standard} (iii). Indeed, the simulation of one trajectory with a time step $10^{-4}$ requires approximately hundred times more than the generation of one trajectory using a time step $10^{-2}$. Hence, a runtime of a few hours would turn to months. In addition, even for \lq\lq arbitrary small\rq\rq \ time steps, one cannot guarantee that the Euler-Maruyama scheme preserves the underlying invariant measure. For these reasons, it is crucial to base our ABC method on the reliable measure-preserving numerical splitting scheme combined with the invariant measure-based distances. 
Our results were discussed in the case of an observable 1-dimensional output process. However, the approach can be directly applied to $d$-dimensional output processes, $d>1$, as long as the underlying SDEs are characterised by an invariant distribution and a measure-preserving numerical method can be derived. In particular,  one can compute the distances in \eqref{7b} for each of the $d$ components and derive a global distance by combining them, e.g., via their sum. Moreover, to account for possible dependences between the observed components,  one can incorporate the cross-spectral densities which are expected to provide further information resulting in an improvement of the performance of the method. An investigation in this direction is currently undergoing. Finally, our proposed ABC method may be also used to investigate invariant manifolds characterised by sets of parameters yielding the same type of data, as illustrated on the stochastic JR-NMM. This may result in a better understanding of the qualitative behaviour of the underlying model and its ability of reproducing the true features of the modelled phenomenon. 

\section{Supplementary Material}
\label{sec:7}
In this supplementary material, we extend the illustration of the performance of the proposed ABC approach by more examples. In particular, we consider two additional SDEs. First, the critically damped harmonic oscillator \eqref{MP}, fulfilling $\lambda^2-\gamma^2=0$, for which an exact simulation of sample paths is possible, allowing for a validation of Algorithm \ref{Algorithm_Standard} (i). Second, a non-linear weakly damped stochastic oscillator, for which we need to apply a measure-preserving numerical splitting scheme, and thus investigate Algorithm \ref{Algorithm_Standard} (ii). Moreover, we also report the simultaneous inference of the new parameters $\theta=(\sigma,\mu)$ in the stochastic JR-NMM \eqref{JR-NMM}, when the connectivity parameter $C$ is known (while in the main manuscript, we estimate $\theta=(\sigma,\mu,C)$). Finally, we report the estimation of $\theta=(\sigma,\mu,C,b)$, based on both simulated and real EEG data.

\subsection{Validation of the Spectral Density-Based ABC Algorithm \ref{Algorithm_Standard} (i)}
\begin{figure}[H]
	\includegraphics[width=1.0\textwidth]{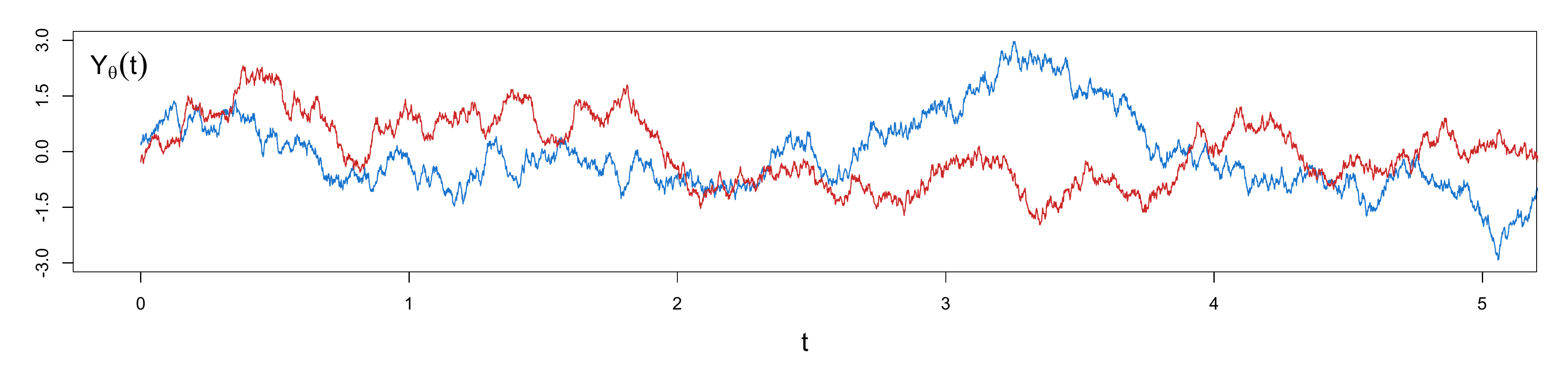}\\
	\caption{Two paths of the output process $\textbf{Y}_\theta=\textbf{P}$ of the critically damped stochastic harmonic oscillator (MP1)  for a noise intensity $\sigma=2$ and parameter $\gamma=1$}
	\label{MP1_path}
\end{figure}

\begin{figure}[H]
	\includegraphics[width=1.0\textwidth]{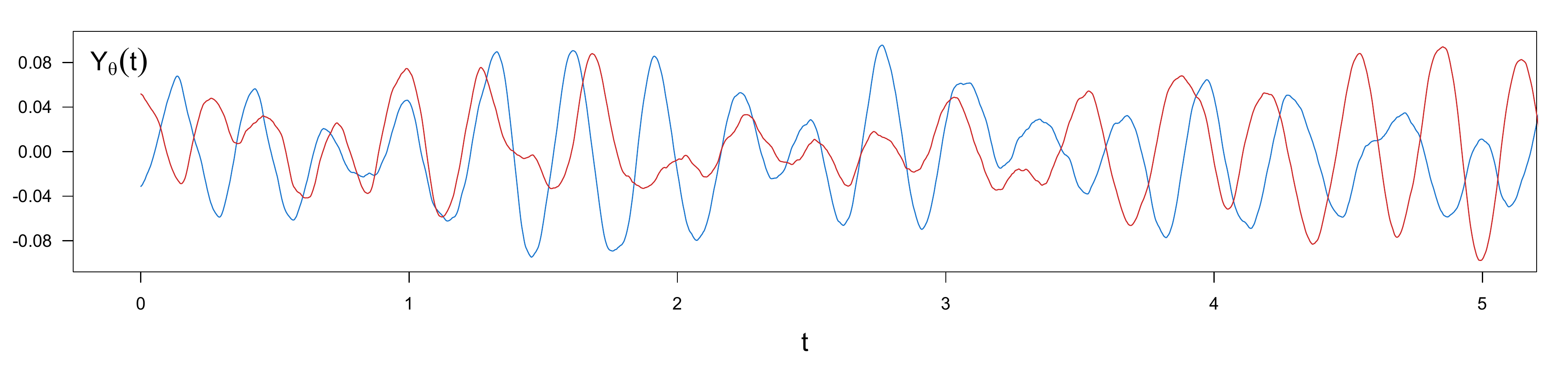}\\
	\caption{Two paths of the output process $\textbf{Y}_\theta=\textbf{Q}$ of the weakly damped stochastic harmonic oscillator (MP2) for a noise intensity $\sigma=2$,  $\gamma=1$ and a damping force $\lambda=20$}
	\label{MP3_path}
\end{figure}

We denote by Model Problem $1$ (MP1) the critically damped harmonic oscillator obtained from \eqref{MP} with $\lambda^2-\gamma^2=0$ (introduced below), and with Model Problem $2$ (MP2) the weakly damped harmonic oscillator, satisfying $\lambda^2-\gamma^2>0$ (see Section \ref{sec:4} of the main manus\-cript). 
Figure \ref{MP1_path} shows two realisations of the output process of MP1 generated with the same choice of parameters. Figure \ref{MP3_path} shows two paths of the output process of MP2 simulated under the same parameter setting.
We perform a step by step investigation of Algorithm \ref{Algorithm_Standard} (i), starting with the estimation of one single model parameter and closing with the successful inference of all parameters.  

\subsubsection{Critically damped stochastic harmonic oscillator: The model and its properties}
We recall the harmonic oscillator \eqref{MP}, focusing on the critically damped case, i.e., $\lambda^2-\gamma^2=0$.   
We assume that the $2$-dimensional process $\textbf{X}=(\textbf{Q},\textbf{P})^{'}$ is partially observed through the second component, i.e., $\textbf{Y}_\theta=\textbf{P}$. 
The invariant distribution $\eta_\textbf{X}$ of the process $\textbf{X}$ is given by
\begin{equation*}
\eta_\textbf{X}=\lim\limits_{t \to \infty} \eta_\textbf{X}(t) =
\mathcal{N}\left(\begin{pmatrix}
0  \\
0  
\end{pmatrix},\begin{pmatrix}
\frac{\sigma^2}{4\gamma^3} & 0 \\
0 & \frac{\sigma^2}{4\gamma} 
\end{pmatrix}\right). 
\end{equation*}
Consequently, the invariant distribution $\eta_{\textbf{Y}_\theta}$ of the output process $\textbf{Y}_\theta$ equals
\begin{equation*}
\eta_{\textbf{Y}_\theta} =
\mathcal{N}\left(0,\frac{\sigma^2}{4\gamma}\right),
\end{equation*}
and the autocovariance function is given by
\begin{equation*}
r_\theta(\Delta) =
\frac{\sigma^2}{4}e^{-\gamma \Delta}\left[\frac{1}{\gamma}-\Delta\right].
\end{equation*}
\subsubsection{Task 1: Inferring one parameter}
At first, we estimate one specific parameter $\theta$ of the model problems, keeping the others fixed. For MP1, we set $\theta=\gamma$ and fix $\sigma=2$. For MP2, we focus on $\theta=\lambda$, fixing $\gamma=1$ and $\sigma=2$. The ABC Algorithm \ref{Algorithm_Standard} (i) is applied to both model problems with $M=10$ observed paths simulated with the exact scheme \eqref{Exact_SDE} using a time step $\Delta =10^{-2}$ over a time interval of length $T=10^3$. In addition, we generate $N=10^5$ synthetic datasets over the same time domain with equal time steps using the exact simulation scheme \eqref{Exact_SDE}. We set the tolerance level to $\epsilon=1^{\text{st}}$ percentile of the calculated distances. Furthermore, we choose uniform prior distributions $\pi(\theta)$ according to
\begin{equation*}
\theta=\left\{\begin{array}{ll} 
\gamma \sim \mathcal{U}(0.1,5)
& \text{for MP1}, \\
\begin{footnotesize}
\textcolor{white}{t}
\end{footnotesize} & \begin{footnotesize}
\textcolor{white}{t}
\end{footnotesize} \\
\lambda \sim \mathcal{U}(10,30) 
& \text{for MP2}
\end{array}\right. .
\end{equation*}
We use the same parameter setting as in Figure \ref{MP1_path} and Figure \ref{MP3_path} for the simulation of the observed reference datasets. In particular, the true parameter values are
\begin{equation*}
\theta^t=\left\{\begin{array}{ll} 
\gamma^t=1 
& \text{for MP1}, \\
\begin{footnotesize}
\textcolor{white}{t}
\end{footnotesize} & \begin{footnotesize}
\textcolor{white}{t}
\end{footnotesize} \\
\lambda^t=20 
& \text{for MP2}
\end{array}\right. .
\end{equation*}

\begin{figure}[H]
	\includegraphics[width=1.0\textwidth]{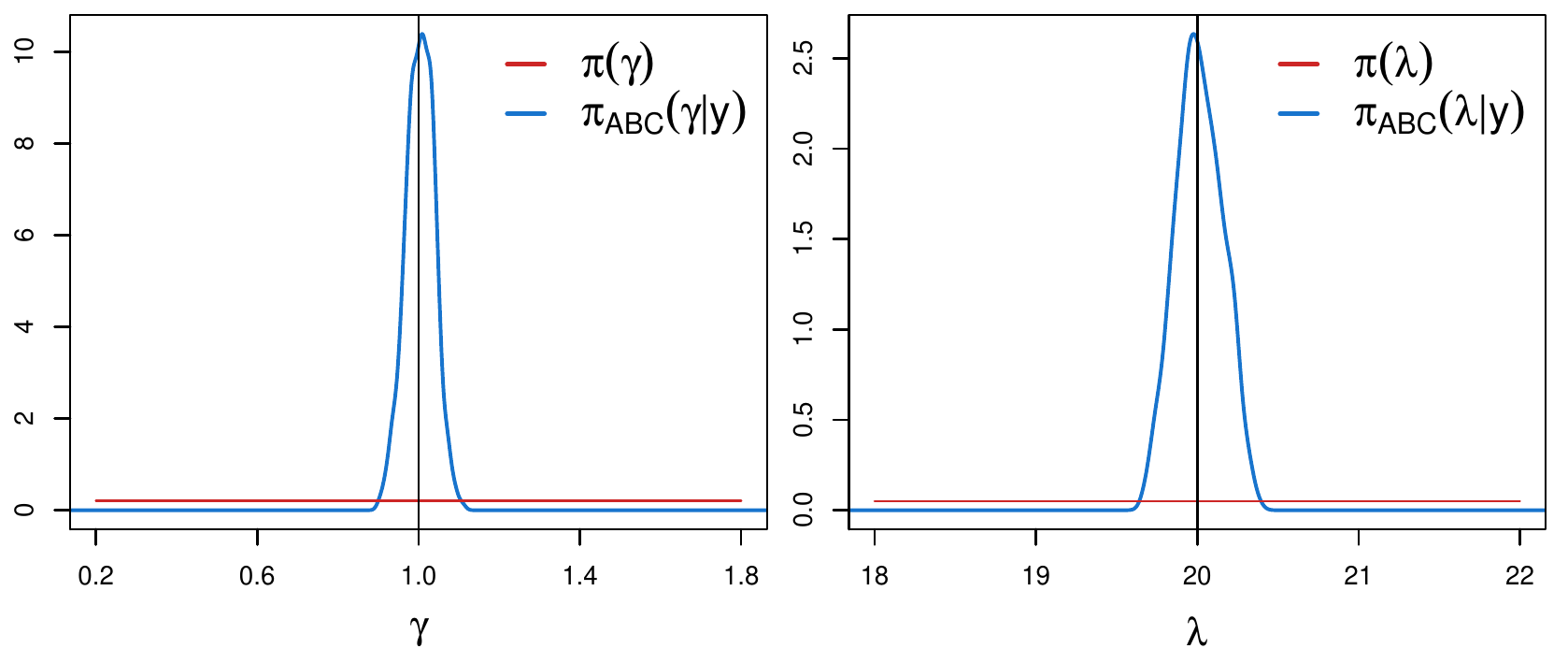}\\
	\caption{ABC posterior densities $\pi_{\text{ABC}}(\theta|y)$ (blue lines) of MP1 (left panel) and MP2 (right panel) obtained from Algorithm \ref{Algorithm_Standard} (i). The horizontal red and vertical black lines denote the uniform priors (not shown according to the full domain) and the true parameter values, respectively}
	\label{1_Par}
\end{figure}

In Figure \ref{1_Par}, we report the results obtained from the proposed Spectral Density-Based ABC Algorithm \ref{Algorithm_Standard} (i). The left panel (referring to MP1) and the right panel (referring to MP2) show the ABC posterior densities $\pi_{\text{ABC}}(\theta|y)$ (blue lines). The horizontal red and vertical black lines denote the prior densities and the true parameter values, respectively. It is remarkable how the flat uniform prior densities are updated by means of the observed data resulting in narrow and unimodal posterior densities that are centered around the true parameter values. 
The ABC posterior means for this and the other scenarios (i.e. the inference of two and three parameters) are reported in Table \ref{table1}.

\subsubsection{Task 2: Inferring two parameters}
We aim for the simultaneous estimation of two parameters, keeping the parameter $\sigma=2$ fixed in MP2. In particular, we consider $\theta=(\gamma,\sigma)$ for MP1 and $\theta=(\lambda,\gamma)$ for MP2. We apply Algorithm \ref{Algorithm_Standard} (i) combined with the exact scheme \eqref{Exact_SDE} under the same values for $M$, $\Delta$ and $T$ as before. Now, we generate $N=5 \cdot 10^5$ synthetic datasets and fix $\epsilon=0.2^{\text{nd}}$ percentile of the calculated distances, keeping the same number of ABC posterior samples as before. We choose the independent uniform priors $\pi(\theta_j)$ according to
\begin{equation*}
\theta_j =\left\{\begin{array}{ll} 
\gamma \sim \mathcal{U}(0.01,2.01) \ \text{and} \ \sigma \sim \mathcal{U}(1,3)
& \text{for MP1}, \\
\begin{footnotesize}
\textcolor{white}{t}
\end{footnotesize} & \begin{footnotesize}
\textcolor{white}{t}
\end{footnotesize} \\
\lambda \sim \mathcal{U}(18,22) \ \text{and} \ \gamma \sim \mathcal{U}(0.01,2.01)
& \text{for MP2}
\end{array}\right. .
\end{equation*}
The true parameter values are
\begin{equation*}
\theta^t=\left\{\begin{array}{ll} 
(\gamma^t,\sigma^t)=(1,2) 
& \text{for MP1}, \\
\begin{footnotesize}
\textcolor{white}{t}
\end{footnotesize} & \begin{footnotesize}
\textcolor{white}{t}
\end{footnotesize} \\
(\lambda^t,\gamma^t)=(20,1) 
& \text{for MP2}
\end{array}\right.. \
\end{equation*}

\begin{figure}[H]
	\centering			
	\subfigure{\includegraphics[width=1.0\textwidth]{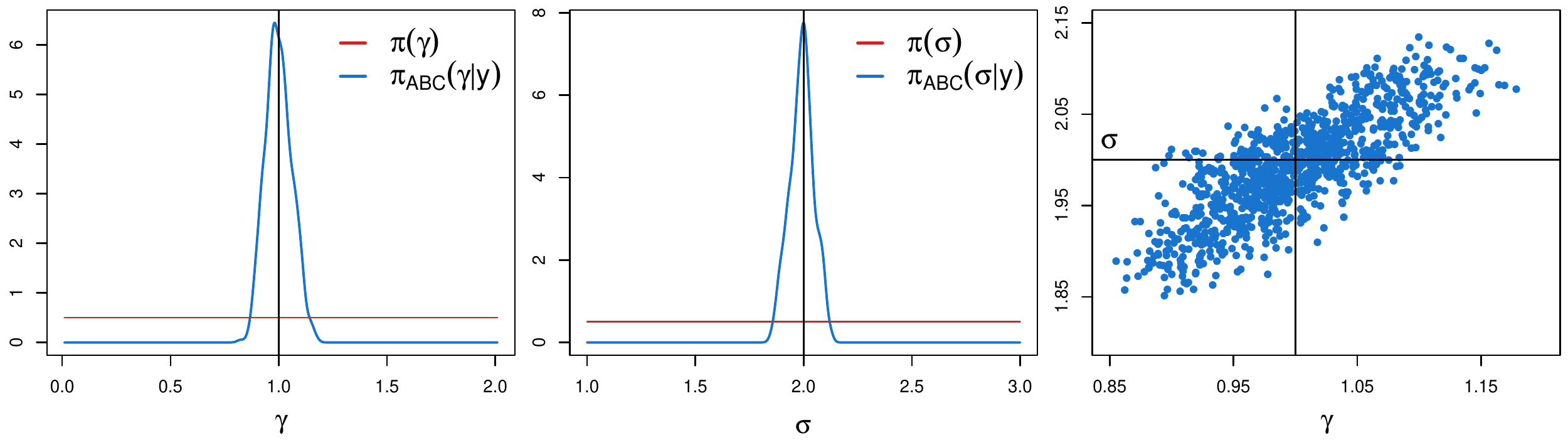}}	
	\subfigure{\includegraphics[width=1.0\textwidth]{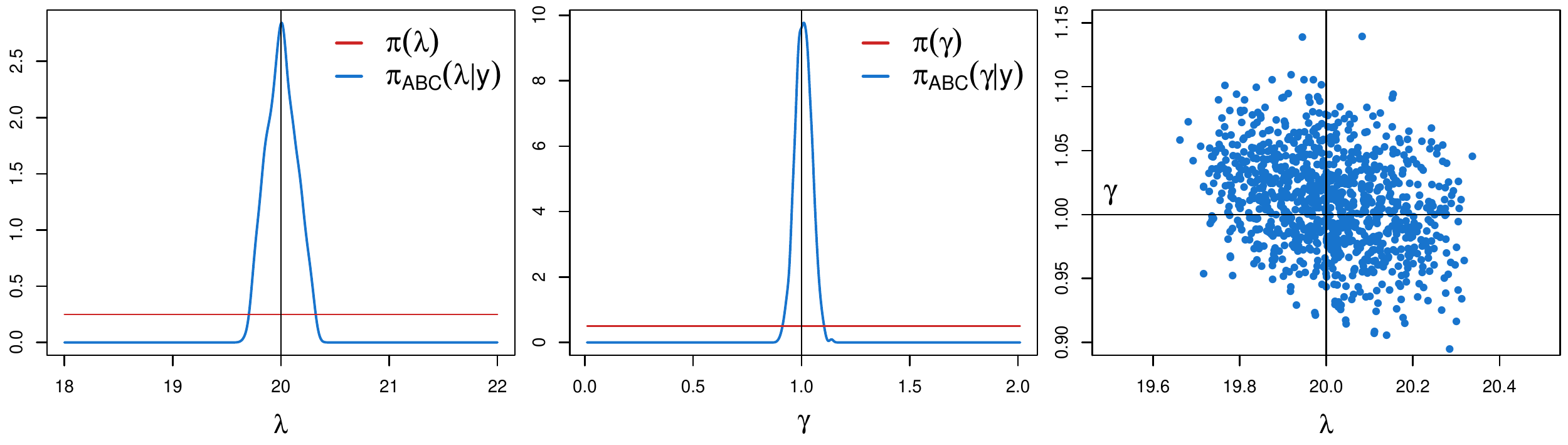}}
	\caption{ABC marginal posterior densities $\pi_{\text{ABC}}(\theta_j|y)$ (blue lines) of MP1 (left and middle top panels) and MP2 (left and middle lower panels) obtained from Algorithm \ref{Algorithm_Standard} (i). The horizontal red and vertical black lines denote the uniform priors and the  true parameter values, respectively. Scatterplots of the kept ABC posterior samples for MP1 and MP2 are reported in the right top and right lower panel, respectively }
	\label{2_Par}
\end{figure}

The ABC marginal posterior densities  $\pi_{\text{ABC}}(\theta_j|y)$ (blue lines) are reported in the left and middle panels of Figure \ref{2_Par} for MP1 (top panels) and MP2 (lower panels),
while the right panels of Figure \ref{2_Par} show the scatterplots of the kept ABC posterior samples. Also in this case, the posteriors are unimodal and centered around the true parameter values. 
Note that, the support of $\pi_{\text{ABC}}(\lambda|y)$ for MP2 is approximately the same as in Figure \ref{1_Par}, suggesting that, in the case of inferring two parameters, the proposed ABC method is able to identify the same region for $\lambda$ as in the case of estimating one parameter. The reason is that the kept ABC posterior samples of $\lambda$ and $\gamma$ are not correlated, as it can be observed in the right lower panel of Figure \ref{2_Par}. On the contrary, the support of $\pi_{\text{ABC}}(\gamma|y)$ for MP1 is broader than in Figure \ref{1_Par}, due to a correlation among the kept ABC posterior samples of $\gamma$ and $\sigma$ (cf. right top panel of Figure \ref{2_Par}). In spite of this, the ABC marginal posterior density resembles that derived when estimating only one parameter (cf. left panel of Figure  \ref{1_Par}). 

\subsubsection{Task 3: Inferring three parameters}
The last goal is the simultaneous inference of all the three parameters $\theta=(\lambda,\gamma,\sigma)$ of MP2.\footnote{Task 3 is already presented in Subsection \ref{subsec:4:2} of the main manuscript. For completeness, we recall it here.} In Figure \ref{3_Par_MP3} (top panels) we report the ABC marginal posterior densities (blue lines) and the prior densities (red lines). In the lower panels, we show the pairwise scatterplots of the kept ABC posterior samples.  The kept posterior values of $\lambda$ turned out to be not correlated with those of the other two parameters, yielding  approximately the same support as in Figure \ref{1_Par} and Figure \ref{2_Par}. Similar to MP1, the kept ABC posterior samples of $\gamma$ and $\sigma$ are correlated (cf. right lower panel of Figure \ref{3_Par_MP3}), leading to a support for $\gamma$ broader than that in Figure \ref{2_Par}. 
The ABC marginal posterior densities shown in Figures \ref{3_Par_MP3}, \ref{1_Par} and \ref{2_Par}, and the results reported in Table \ref{table1} highlight the good performance of the proposed Spectral Density-Based ABC Algorithm \ref{Algorithm_Standard} (i) under the optimal condition of exact, and thus measure-preservative data simulation from the underlying model.
\begin{table}[H]       
	\caption{Parameters of interest, true parameter values and ABC posterior means}
	\begin{center}
		\label{table1}
		\begin{tabular}{lll}
			\hline\noalign{\smallskip}
			$\theta$ & $\theta^t$ & $\hat{\theta}_{\textrm{ABC}}$ \\
			\noalign{\smallskip}\hline\noalign{\smallskip}
			\textcolor{blue}{MP1} & & \\
			$\gamma$ & $1$ & $1.004$ \\
			$(\gamma,\sigma)$ & $(1,2)$ & $(0.9995,1.991)$ \\
			\textcolor{blue}{MP2} & & \\
			$\lambda$ & $20$ & $20.014$ \\
			$(\lambda,\gamma)$ & $(20,1)$ & $(20.005,1.009)$ \\
			$(\lambda,\gamma,\sigma)$ & $(20,1,2)$ & $(20.015,1.002,2.011)$ \\
			\noalign{\smallskip}\hline
		\end{tabular}
	\end{center}
\end{table}

\subsection{Validation of the Spectral Density-Based and Measure-Preserving ABC Algorithm \ref{Algorithm_Standard} (ii) on an extended version of MP2}
We now consider an extended non-linear version of the previously studied Model Problem $2$. Due to the non-linearity in the model, an exact simulation scheme is not available. Hence, we consider the measure-preserving numerical splitting scheme \eqref{Strang}, and thus investigate the performance of Algorithm \ref{Algorithm_Standard} (ii).
\subsubsection{A non-linear weakly damped stochastic oscillator}
\begin{figure}[H]
	\includegraphics[width=1.0\textwidth]{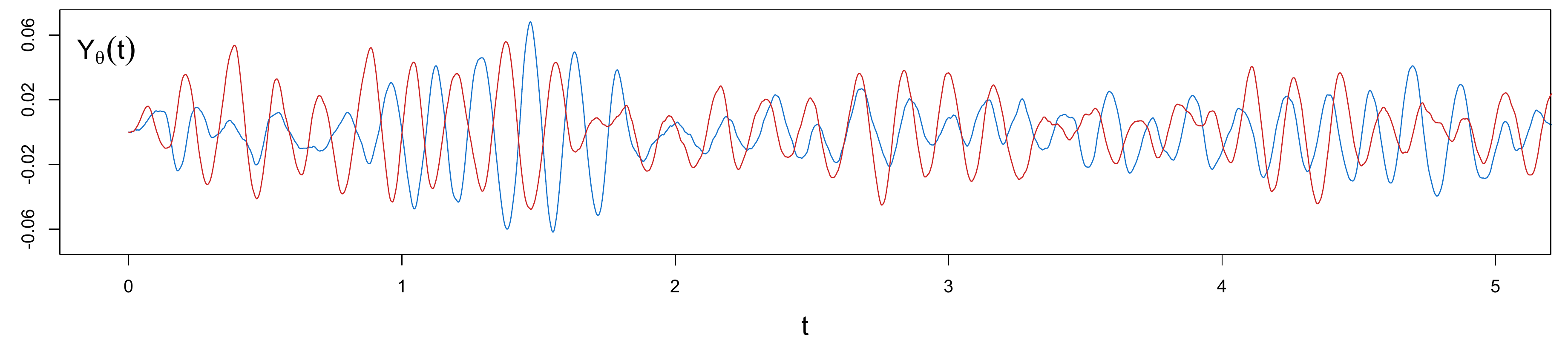}\\
	\caption{Two paths of the output process $\textbf{Y}_\theta=\textbf{Q}$ of the non-linear stochastic oscillator \eqref{MP4} for $\theta=(\lambda,\gamma,\sigma)=(20,1,2)$} \label{Path_MP4}
\end{figure}

We consider a stochastic oscillator that incorporates a high-amplitude sine wave represented by the non-linear displacement term $G(\textbf{Q})=-10^3 \sin(\textbf{Q})$. In particular, we study the $2$-dimensional SDE
\begin{equation}\label{MP4}
d \begin{pmatrix}
Q(t) \\
P(t) 
\end{pmatrix}
=\begin{pmatrix}
P(t) \\
-\lambda^2Q(t) -2\gamma P(t) + G(Q(t))
\end{pmatrix} dt + \begin{pmatrix}
0 \\
\sigma
\end{pmatrix} dW(t)
\end{equation}
with the strictly positive parameters $\theta=(\lambda,\gamma,\sigma)$. The condition $\lambda^2-\gamma^2>0$ guarantees a weak damping. 
The $2$-dimensional solution process $\textbf{X}=(\textbf{Q},\textbf{P})^{'}$ is partially observed through the first coordinate, i.e., $\textbf{Y}_\theta=\textbf{Q}$. Figure \ref{Path_MP4} shows two realisations of the output process generated with the same choice of parameters.

\subsubsection{Parameter inference from simulated data}
We assume to observe $M=30$ paths of the output process simulated with the measure-preserving numerical scheme \eqref{Strang} 
over a time interval of length $T=10^3$ using a time step $\Delta=10^{-2}$ and the same true parameter values as in Figure \ref{Path_MP4}, i.e.,
\begin{equation*}
\theta^t=(\lambda^t,\gamma^t,\sigma^t)=(20,1,2).
\end{equation*}
We then use the same $T$ and $\Delta$ to generate $N=2 \cdot 10^6$ synthetic datasets within ABC. We further choose the tolerance level $\epsilon=0.05^{\text{th}}$ percentile of the calculated distances, set $w=0$ in \eqref{weight} and use independent uniform prior distributions 
\begin{equation*}
\lambda \sim \mathcal{U}(18,22), \quad \gamma \sim \mathcal{U}(0.01,2.01) \quad \text{and} \quad \sigma \sim \mathcal{U}(1,3).
\end{equation*}

\begin{figure}[H]
	\includegraphics[width=1.0\textwidth]{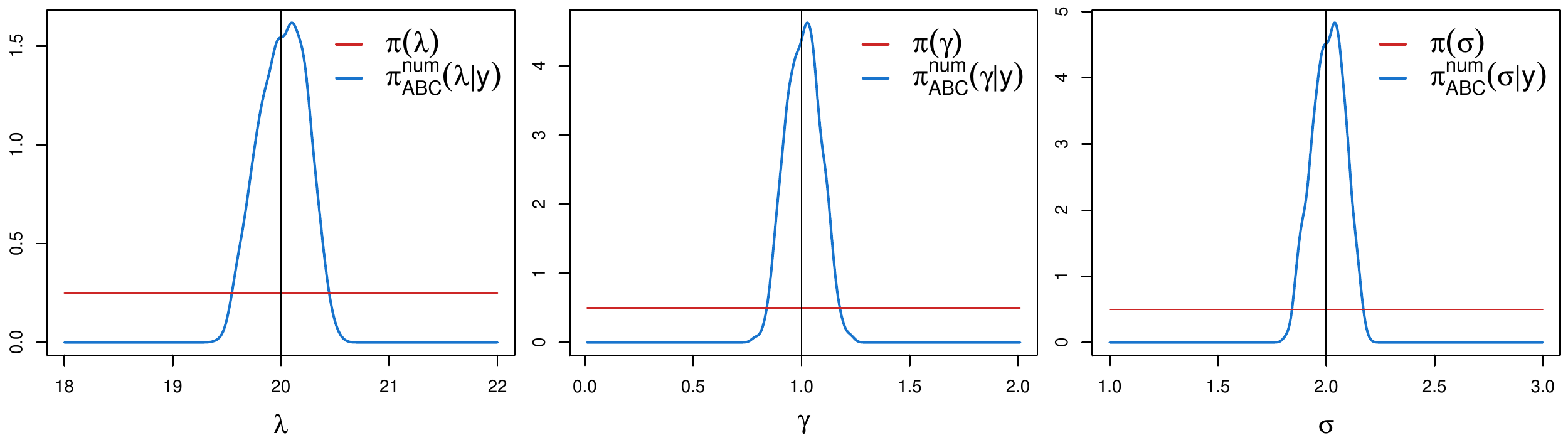}\\
	\caption{ABC marginal posterior densities $\pi_{\text{ABC}}^{\text{num}}(\theta_j|y)$ (blue lines) of $\theta=(\lambda,\gamma,\sigma)$ of the non-linear weakly damped stochastic oscillator \eqref{MP4} obtained from Algorithm \ref{Algorithm_Standard} (ii). The horizontal red and vertical black lines denote the uniform priors and the true parameter values, respectively} \label{results_MP4} 
\end{figure}

Figure \ref{results_MP4} shows the  ABC marginal posterior densities $\pi_{\text{ABC}}^{\textrm{num}}(\theta_j|y)$. They are unimodal, narrow and centered around the true parameter values. 
The ABC posterior means are given by
\begin{equation*}
(\hat{\lambda}_{\textrm{ABC}}^{\text{num}},\hat{\gamma}_{ABC}^{\text{num}},\hat{\sigma}_{ABC}^{\text{num}})=(20.015,1.008,2.0105).
\end{equation*}
In spite of the presence of the non-linear term $G$, the inference via Algorithm \ref{Algorithm_Standard} (ii) yields  results similar to those obtained for MP2 when applying Algorithm \ref{Algorithm_Standard} (i) under the exact data generation. 

\subsection{Application of the Spectral Density-Based and Measure-Preserving ABC Algorithm \ref{Algorithm_Standard} (ii) for the inference of the new parameters $\theta=(\sigma,\mu)$ of the stochastic JR-NMM}
We now estimate $\theta=(\sigma,\mu)$ of the stochastic JR-NMM \eqref{JR-NMM} (see Section \ref{sec:5} of the main manuscript). These are new parameters introduced by \cite{2} in the SDE reformulation of the original JR-NMM \citep{1}. Differently from the other parameters, these parameters have not yet been estimated in the literature. Here, we fix $C=135$ and apply Algorithm \ref{Algorithm_Standard} (ii) with $M=30$, $N=5\cdot 10^5$, $\Delta=2\cdot 10^{-3}$ and $T=200$. We fix $\epsilon=0.2^{\text{nd}}$ percentile of the calculated distances and choose  uniform prior distributions according to
\begin{equation*}
\sigma \sim \mathcal{U}(1300,2700) \quad \text{and} \quad \mu \sim \mathcal{U}(160,280).
\end{equation*}
The true parameter values used to generate the observed data are given by
\begin{equation*}
\theta^t=(\sigma^t,\mu^t)=(2000,220).
\end{equation*}

\begin{figure}[H]
	\centering			
	\subfigure{\includegraphics[width=1.0\textwidth]{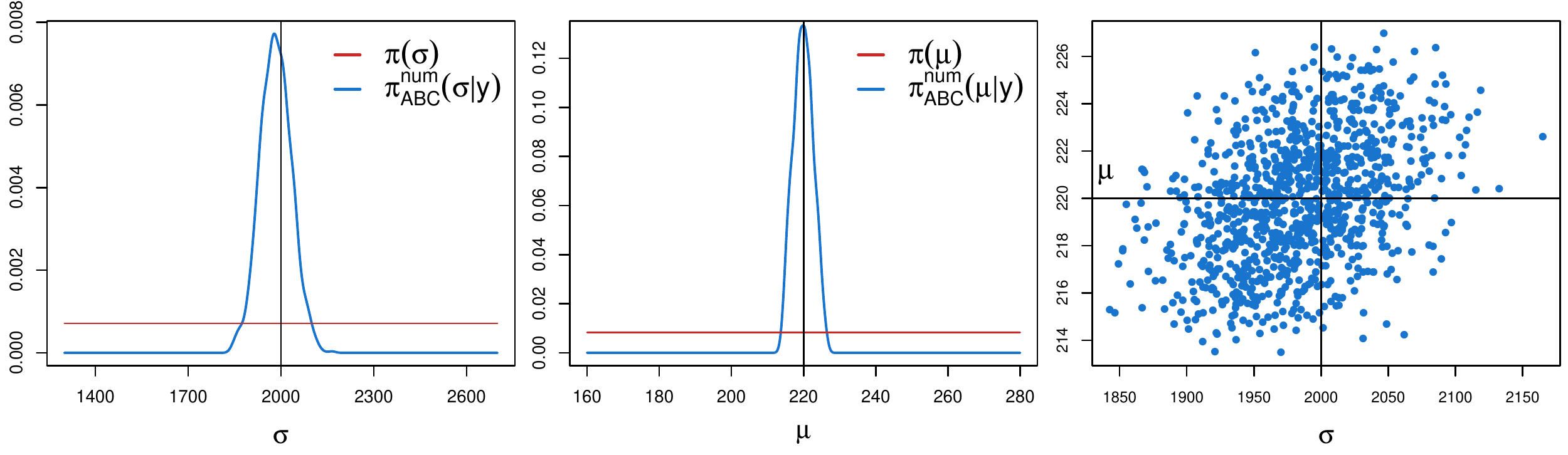}}	
	\subfigure{\includegraphics[width=0.6\textwidth]{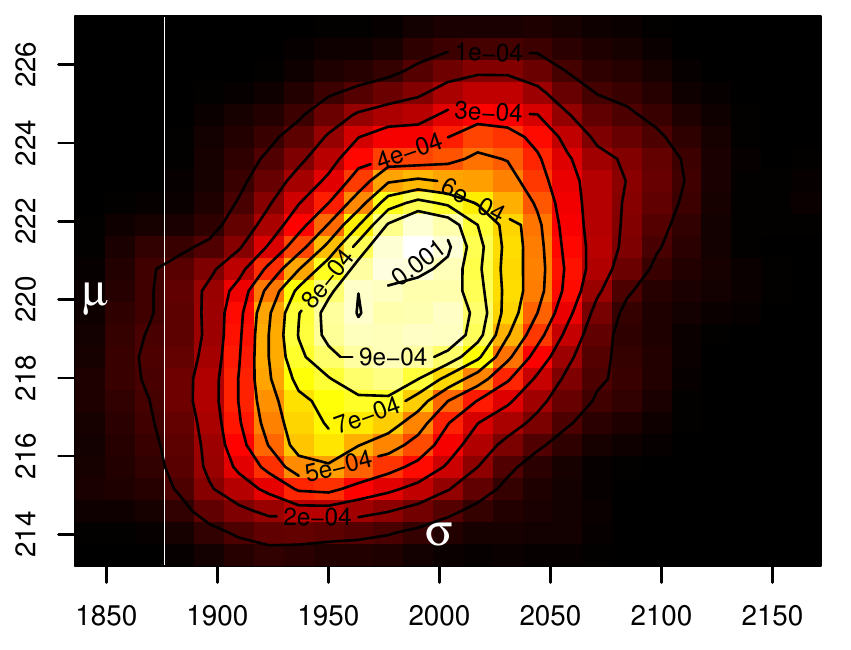}}
	\caption{ABC marginal posterior densities $\pi_{\text{ABC}}^{\text{num}}(\theta_j|y)$ (blue lines, left and middle top panels) 
		of $\theta=(\sigma,\mu)$ of the stochastic JR-NMM \eqref{JR-NMM} obtained from Algorithm \ref{Algorithm_Standard} (ii). The horizontal red and vertical black lines denote the uniform priors and the true parameter values, respectively. Scatterplot of the kept ABC posterior samples (right top panel) and contour plot of the ABC posterior density (lower panel)} 
	\label{JRNMM_sigmu}
\end{figure}

Figure \ref{JRNMM_sigmu} shows the  ABC marginal posterior densities $\pi_{\textrm{ABC}}^{\textrm{num}}(\theta_j|y)$ (left and middle top panels) obtained by applying the Spectral Density-Based and Measure-Preserving ABC Algorithm \ref{Algorithm_Standard} (ii). The posteriors are  centered around the true parameter values, leading to marginal ABC posterior means given by
\begin{equation*}
(\hat{\sigma}_{\textrm{ABC}}^\textrm{num},\hat{\mu}_{\textrm{ABC}}^\textrm{num})=(1985.936,220.1364).
\end{equation*}
From the scatterplot of the kept ABC posterior samples of $\sigma$ and $\mu$ (right top panel),  we conclude that they are not correlated. The successful performance of the proposed ABC approach is also visible by looking at the contour plot of the ABC posterior density (lower panel). Indeed, the proposed algorithm is able to detect a plain region of posterior values for $\theta$ around $\theta^t$.

\subsection{Application of the Spectral Density-Based and Measure-Preserving ABC Algorithm \ref{Algorithm_Standard} (ii) for the inference of $\theta=(\sigma,\mu,C,b)$ of the stochastic JR-NMM}
We now demonstrate that we obtain satisfactory results even when inferring the four parameters $\theta=(\sigma,\mu,C,b)$ of the stochastic JR-NMM \eqref{JR-NMM}. Since the parameters of main interest are $\sigma$, $\mu$ and $C$, in the main manuscript (see Section $5$) we did not take into account the well-reported coefficient $b$, which takes the value $b=50$ in the literature; see, e.g., \cite{1} and the references therein.
\subsubsection{Inference from simulated data}
We start with inferring $\theta=(\sigma,\mu,C,b)$ from simulated data and
apply Algorithm \ref{Algorithm_Standard} (ii) for $M=30$, $N=5\cdot 10^6$, $\Delta=2\cdot 10^{-3}$ and $T=200$. We fix $\epsilon=0.004^{\text{th}}$ percentile and use the following uniform priors
\begin{equation*}
\sigma \sim \mathcal{U}(1300,2700), \quad \mu \sim \mathcal{U}(160,280), 
\end{equation*}
\begin{equation*}
C \sim \mathcal{U}(129,141) \quad \text{and} \quad b \sim \mathcal{U}(44,56).
\end{equation*}
The reference data is generated under
\begin{equation*}
\theta^t=(\sigma^t,\mu^t,C^t,b^t)=(2000,220,135,50).
\end{equation*}

\begin{figure}[H]
	\centering			
	\subfigure{\includegraphics[width=1.0\textwidth]{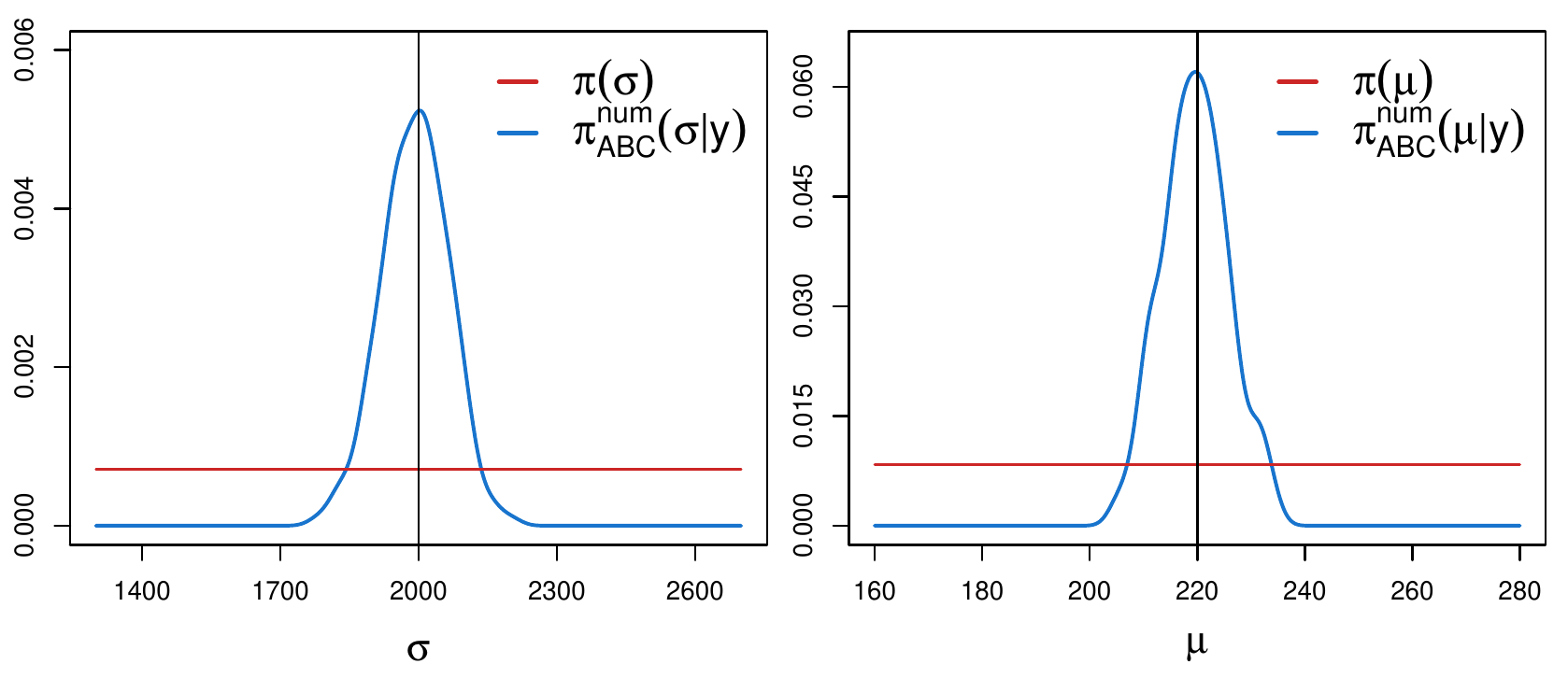}}	
	\subfigure{\includegraphics[width=1.0\textwidth]{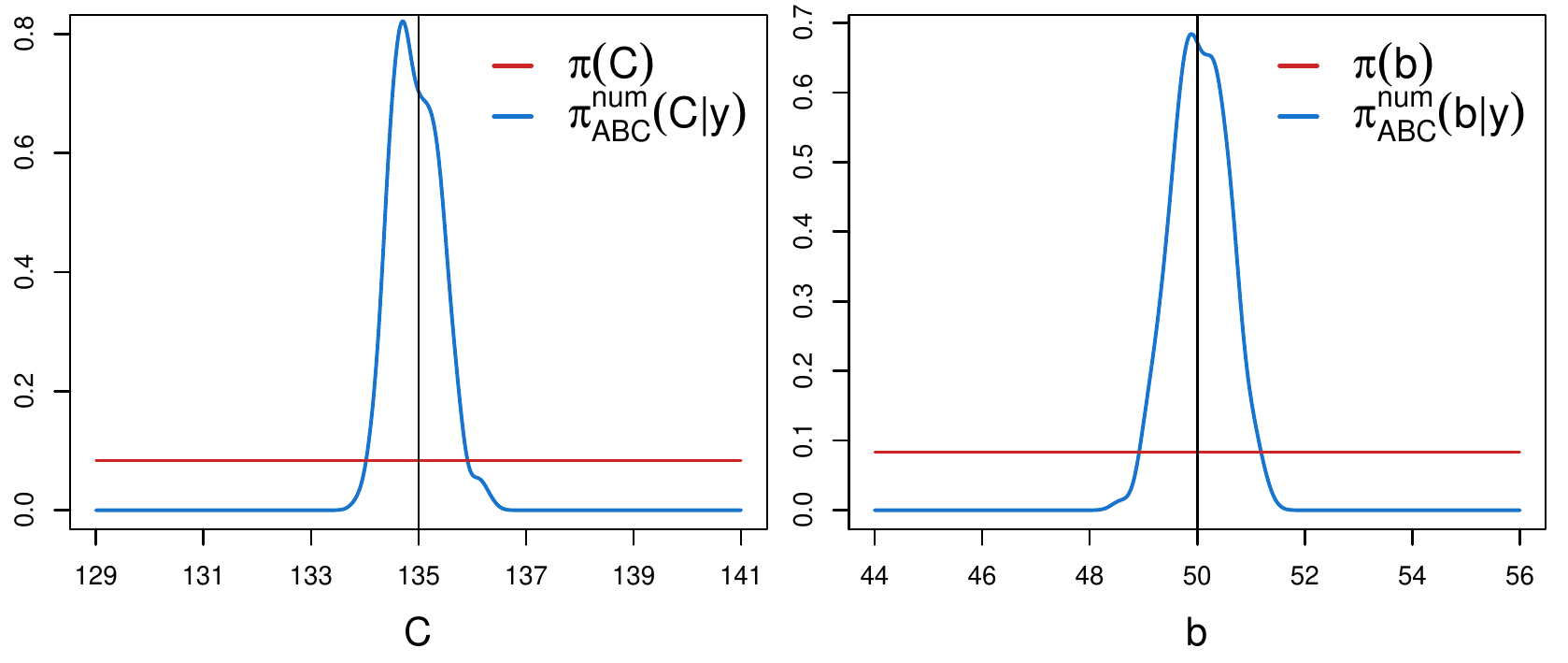}}
	\caption{ABC marginal posterior densities $\pi_{\text{ABC}}^{\text{num}}(\theta_j|y)$ (blue lines) of $\theta=(\sigma,\mu,C,b)$ of the stochastic JR-NMM \eqref{JR-NMM} obtained from Algorithm \eqref{Algorithm_Standard} (ii). The horizontal red lines and the vertical black lines represent the uniform priors and the true parameter values, respectively} 
	\label{JRNMM_sigmucb}
\end{figure}

In Figure \ref{JRNMM_sigmucb}, we report the marginal ABC posterior densities $\pi_{\textrm{ABC}}^{\textrm{num}}(\theta_j|y)$, which are again centered around the true parameter values. The marginal posterior means are given by 
\begin{equation*}
(\hat{\sigma}_{\textrm{ABC}}^\textrm{num},\hat{\mu}_{\textrm{ABC}}^\textrm{num},\hat{C}_{\textrm{ABC}}^\textrm{num},\hat{b}_{\textrm{ABC}}^\textrm{num}) 
=(1992.6,219.7,134.95,50.05).
\end{equation*}

\subsubsection{Inference from real EEG data}
Finally, we infer $\theta=(\sigma,\mu,C,b)$ from real EEG data. Algorithm \ref{Algorithm_Standard} (ii) is applied under the same conditions as in Subsection \ref{subsec:5:3} of the main manuscript, except fixing $\epsilon=0.002^{\text{nd}}$ percentile of calculated distances and choosing the uniform priors according to
\begin{equation*}
\sigma \sim \mathcal{U}(500,3500), \quad \mu \sim \mathcal{U}(70,370), 
\end{equation*}
\begin{equation*}
C \sim \mathcal{U}(120,150) \quad \text{and} \quad b \sim \mathcal{U}(40,60).
\end{equation*}
\begin{figure}[H]
	\centering			
	\subfigure{\includegraphics[width=1.0\textwidth]{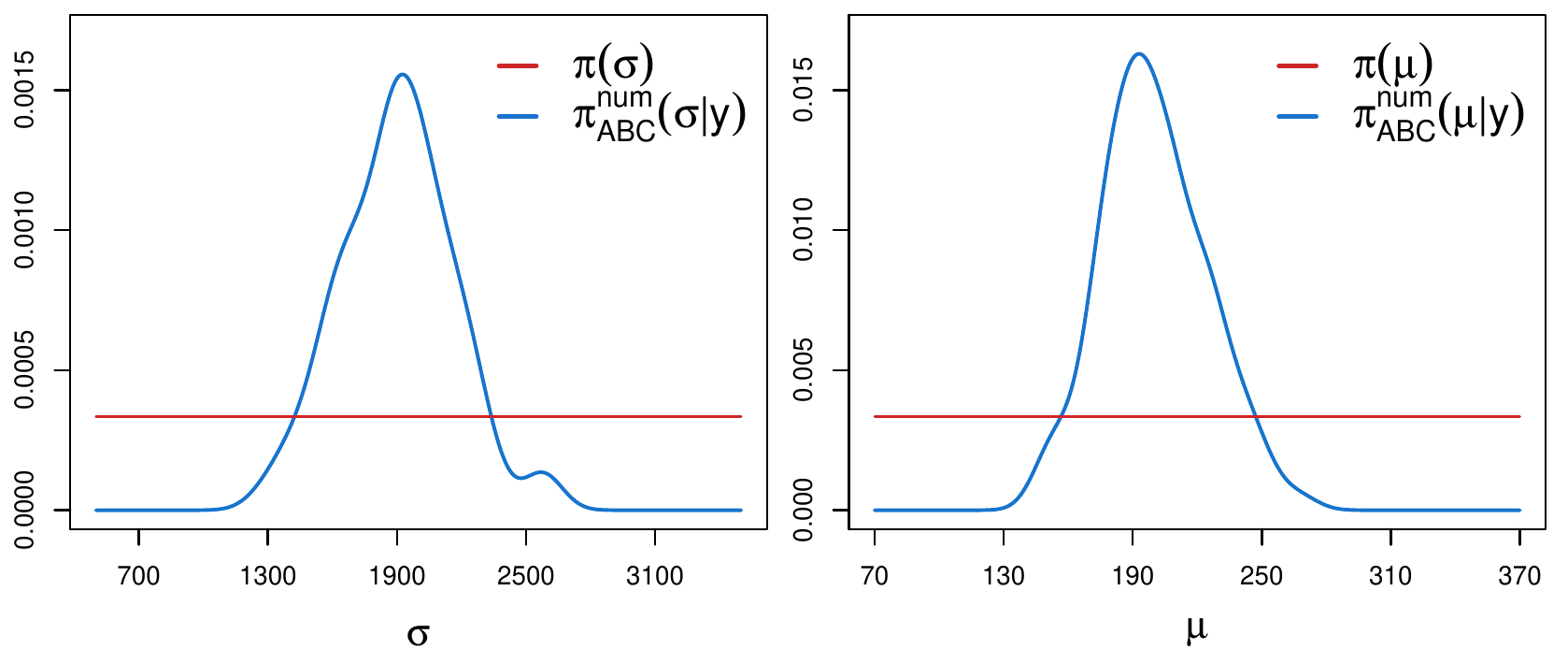}}	
	\subfigure{\includegraphics[width=1.0\textwidth]{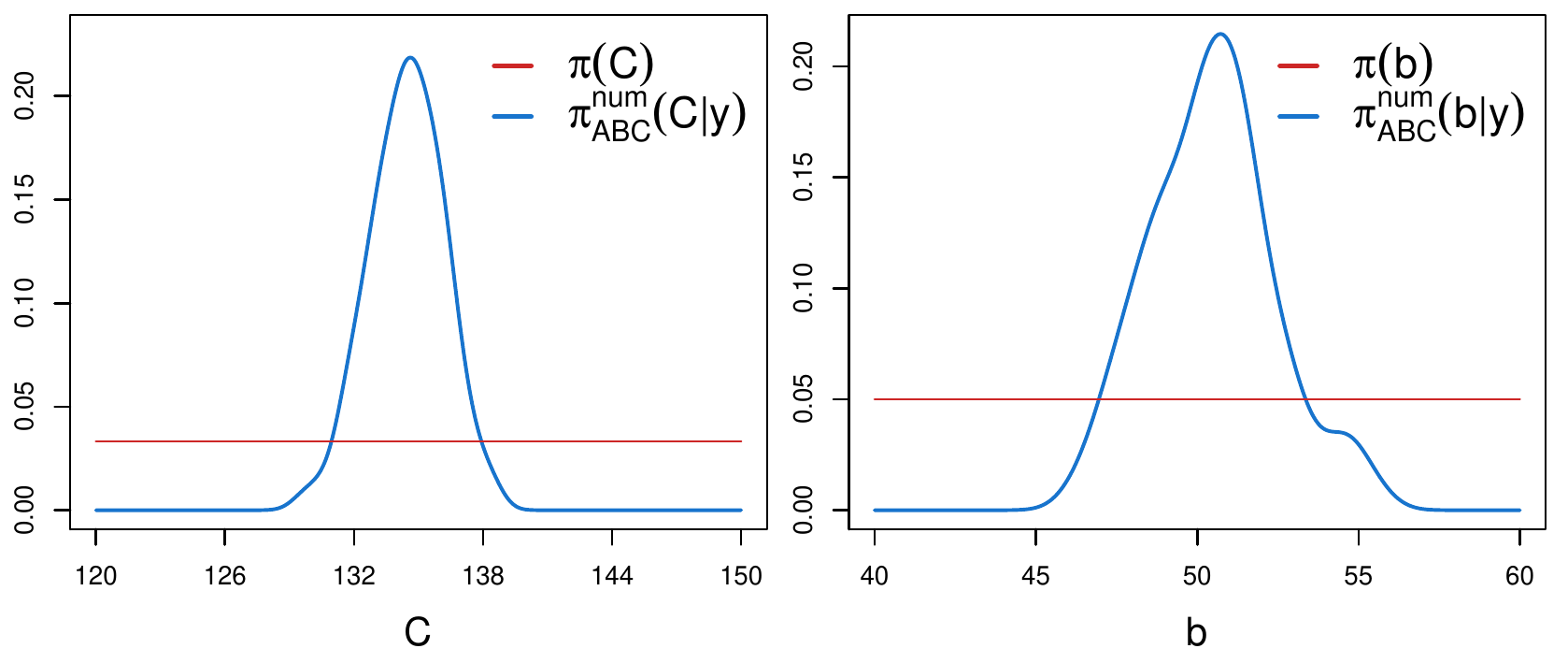}}
	\caption{Marginal ABC posterior densities $\pi_{\text{ABC}}^{\text{num}}(\theta_j|y)$ (blue lines) of $\theta=(\sigma,\mu,C,b)$ of the stochastic JR-NMM \eqref{JR-NMM} fitted on real EEG data using Algorithm \eqref{Algorithm_Standard} (ii). The red lines correspond to the uniform priors} 
	\label{EEG_sigmucb}
\end{figure}

Figure \ref{EEG_sigmucb} shows the unimodal marginal ABC posterior densities $\pi_{\textrm{ABC}}^{\textrm{num}}(\theta_j|y)$, yielding posterior means given by
\begin{align*}
\begin{split}
&(\hat{\sigma}_{\textrm{ABC}}^\textrm{num},\hat{\mu}_{\textrm{ABC}}^\textrm{num},\hat{C}_{\textrm{ABC}}^\textrm{num},\hat{b}_{\textrm{ABC}}^\textrm{num}) \\
&=(1902.6,200.2,134.45,50.46). 
\end{split}
\end{align*}
Focusing on the coefficient $b$, the corresponding marginal posterior density is centered around $b=50$, which is the value reported in the literature.

\newpage

\begin{thebibliography}{63}
	\providecommand{\natexlab}[1]{#1}
	\providecommand{\url}[1]{{#1}}
	\providecommand{\urlprefix}{URL }
	\expandafter\ifx\csname urlstyle\endcsname\relax
	\providecommand{\doi}[1]{DOI~\discretionary{}{}{}#1}\else
	\providecommand{\doi}{DOI~\discretionary{}{}{}\begingroup
		\urlstyle{rm}\Url}\fi
	\providecommand{\eprint}[2][]{\url{#2}}
	
	\bibitem[{Ableidinger and Buckwar(2016)}]{Ableidinger2016}
	Ableidinger, M., Buckwar, E.: Splitting Integrators for the Stochastic
	Landau--Lifshitz Equation. SIAM J. Sci. Comput. 38, \penalty0 A1788--A1806
	(2016)
	
	\bibitem[{Ableidinger et~al.(2017)Ableidinger, Buckwar, and Hinterleitner}]{2}
	Ableidinger, M., Buckwar, E., Hinterleitner, H.: A Stochastic Version of the
	Jansen and Rit Neural Mass Model: Analysis and Numerics. J. Math. Neurosci.
	7\penalty0 (8) (2017)
	
	\bibitem[{Andrzejak et~al.(2001)Andrzejak, Lehnertz, Mormann, Rieke, David, and
		Elger}]{7}
	Andrzejak, R.~G., Lehnertz, K., Mormann, F., Rieke, C., David, P., Elger,
	C.~E.: Indications of nonlinear deterministic and finite-dimensional
	structures in time series of brain electrical activity: Dependence on
	recording region and brain state. Phys. Rev. E 64, \penalty0 061907 (2001)
	
	\bibitem[{Arnold(1974)}]{9}
	Arnold, L.: Stochastic differential equations: theory and applications. Wiley,
	New York (1974)
	
	\bibitem[{Barber et~al.(2015)Barber, Voss, and Webster}]{22}
	Barber, S., Voss, J., Webster, M.: The rate of convergence for approximate
	Bayesian computation. Electron. J. Stat. 9\penalty0 (1), \penalty0 80--105
	(2015)
	
	\bibitem[{Barnes et~al.(2012)Barnes, Filippi, Stumpf, and Thorne}]{37}
	Barnes, C., Filippi, S., Stumpf, M., Thorne, T.: Considerate approaches to
	constructing summary statistics for ABC model selection. Stat. Comput.
	22\penalty0 (6), \penalty0 1181--1197 (2012)
	
	\bibitem[{Beaumont et~al.(2002)Beaumont, Zhang, and Balding}]{34}
	Beaumont, M.~A., Zhang, W., Balding, D.~J.: Approximate Bayesian Computation in
	Population Genetics. Genetics 162\penalty0 (4), \penalty0 2025--2035 (2002)
	
	\bibitem[{Bernton et~al.(2019)Bernton, Jacob, Gerber, and Robert}]{10}
	Bernton, E., Jacob, P.~E., Gerber, M., Robert, C.~P.: Approximate Bayesian
	computation with the Wasserstein distance. J. Roy. Stat. Soc. B  (2019)
	
	\bibitem[{Biau et~al.(2015)Biau, C{\'e}rou, and Guyader}]{20}
	Biau, G., C{\'e}rou, F., Guyader, A.: New Insights Into Approximate Bayesian
	Computation. Ann. I. H. Poincare B 51\penalty0 (1), \penalty0 376--403 (2015)
	
	\bibitem[{Blanes et~al.(2009)Blanes, Casas, and Murua}]{Blanes2009}
	Blanes, S., Casas, F., Murua, A.: Splitting and composition methods in the
	numerical integration of differential equations. Bol. Soc. Esp. Mat. Apl. 45
	(2009)
	
	\bibitem[{Blum(2010{\natexlab{a}})}]{21}
	Blum, M.~G.~B.: Approximate Bayesian Computation: A Nonparametric Perspective.
	J. Am. Stat. Assoc. 105\penalty0 (491), \penalty0 1178--1187
	(2010{\natexlab{a}})
	
	\bibitem[{Blum(2010{\natexlab{b}})}]{33}
	Blum, M.~G.~B.: Choosing the Summary Statistics and the Acceptance Rate in
	Approximate Bayesian Computation. In: Lechevallier, Y., Saporta, G. (eds)
	Proceedings of COMPSTAT 2010: Physica-Verlag HD, Heidelberg: pp 47--56
	(2010{\natexlab{b}})
	
	\bibitem[{Boys et~al.(2008)Boys, Wilkinson, and Kirkwood}]{Boysetal2008}
	Boys, R.~J., Wilkinson, D.~J., Kirkwood, T.~B.~L.: Bayesian inference for a
	discretely observed stochastic kinetic model. Stat. Comput. 18, \penalty0
	125--135 (2008)
	
	\bibitem[{Br\'{e}hier and Gouden\`{e}ge(2019)}]{Brehier2018}
	Br\'{e}hier, C.~E., Gouden\`{e}ge, L.: Analysis of Some Splitting Schemes for
	the Stochastic Allen-Cahn Equation. Discrete Cont. Dyn.-B 24, \penalty0
	4169--4190 (2019)
	
	\bibitem[{Cadonna et~al.(2017)Cadonna, Kottas, and Prado}]{61}
	Cadonna, A., Kottas, A., Prado, R.: Bayesian mixture modeling for spectral
	density estimation. Stat. Prob. Lett. 125, \penalty0 189--195 (2017)
	
	\bibitem[{Drovandi et~al.(2016)Drovandi, Pettitt, and McCutchan}]{Drovandietal}
	Drovandi, C.~C., Pettitt, A.~N., McCutchan, R.: Exact and Approximate Bayesian
	Inference for Low Integer-Valued Time Series Models with Intractable
	Likelihoods. Bayesian Anal. 11, \penalty0 325--352 (2016)
	
	\bibitem[{Eddelbuettel and Fran\c{c}ois(2011)}]{Rcpp}
	Eddelbuettel, D., Fran\c{c}ois, R.: {Rcpp}: Seamless {R} and {C++} Integration.
	J. Stat. Soft. 40\penalty0 (8), \penalty0 1--18 (2011)
	
	\bibitem[{Fan and Sisson(2018)}]{FanSisson2018}
	Fan, Y., Sisson, S.~A. (2018) ABC samplers. In: Sisson, S.~A., Fan, Y.,
	Beaumont, M. (eds) Handbook of Approximate Bayesian Computation: CRC Press,
	Taylor \& Francis Group: chap~4, pp 87--123
	
	\bibitem[{Fearnhead and Prangle(2012)}]{38}
	Fearnhead, P., Prangle, D.: Constructing summary statistics for approximate
	Bayesian computation: semi-automatic approximate Bayesian computation. J.
	Roy. Stat. Soc. B 74\penalty0 (3), \penalty0 419--474 (2012)
	
	\bibitem[{Jansen and Rit(1995)}]{1}
	Jansen, B.~H., Rit, V.~G.: Electroencephalogram and visual evoked potential
	generation in a mathematical model of coupled cortical columns. Biol. Cybern.
	73\penalty0 (4), \penalty0 357--366 (1995)
	
	\bibitem[{Jansen et~al.(1993)Jansen, Zouridakis, and Brandt}]{Jansen1993}
	Jansen, B.~H., Zouridakis, G., Brandt, M.~E.: A neurophysiologically-based
	mathematical model of flash visual evoked potentials. Biol. Cybern. 68,
	\penalty0 275--283 (1993)
	
	\bibitem[{Jasra(2015)}]{41}
	Jasra, A.: Approximate Bayesian Computation for a Class of Time Series Models.
	Int. Stat. Rev. 83\penalty0 (3), \penalty0 405--435 (2015)
	
	\bibitem[{Jiang et~al.(2017)Jiang, Wu, Zheng, and Wong}]{Jiang2017}
	Jiang, B., Wu, T.-y., Zheng, C., Wong, W.~H.: Learning summary statistics for
	Approximate Bayesian Computation via deep neural network. Stat. Sinica
	27\penalty0 (4), \penalty0 1595--1618 (2017)
	
	\bibitem[{Kloeden and Platen(1992)}]{REF_I3}
	Kloeden, P.~E., Platen, E.: Numerical Solution of Stochastic Differential
	Equations. Springer, Berlin (1992)
	
	\bibitem[{Kypraios et~al.(2017)Kypraios, Neal, and Prangle}]{52}
	Kypraios, T., Neal, P., Prangle, D.: A tutorial introduction to Bayesian
	inference for stochastic epidemic models using Approximate Bayesian
	Computation. Math. Biosci. 287, \penalty0 42--53 (2017)
	
	\bibitem[{Leimkuhler and Matthews(2015)}]{6}
	Leimkuhler, B., Matthews, C.: Molecular dynamics: with deterministic and
	stochastic numerical methods. Springer International Publ., Cham (2015)
	
	\bibitem[{Leimkuhler et~al.(2016)Leimkuhler, Matthews, and Stoltz}]{4}
	Leimkuhler, B., Matthews, C., Stoltz, G.: The computation of averages from
	equilibrium and nonequilibrium Langevin molecular dynamics. IMA. J. Numer.
	Anal. 36\penalty0 (1), \penalty0 16--79 (2016)
	
	\bibitem[{Lintusaari et~al.(2017)Lintusaari, Gutmann, Dutta, Kaski, and
		Corander}]{12}
	Lintusaari, J., Gutmann, M., Dutta, R., Kaski, S., Corander, J.: Fundamentals
	and Recent Developments in Approximate Bayesian Computation. Syst. Biol.
	66\penalty0 (1), \penalty0 e66--e82 (2017)
	
	\bibitem[{Malham and Wiese(2013)}]{REF13_1}
	Malham, S.~J., Wiese, A.: Chi-square simulation of the CIR process and the
	Heston model. Int. J. Theor. Appl. Finance 16\penalty0 (3) (2013)
	
	\bibitem[{Marin et~al.(2012)Marin, Pudlo, Robert, and Ryder}]{25}
	Marin, J.-M., Pudlo, P., Robert, C.~P., Ryder, R.: Approximate Bayesian
	computational methods. Stat. Comput. 22\penalty0 (6), \penalty0 1167--1180
	(2012)
	
	\bibitem[{Martin et~al.(2019)Martin, McCabe, Frazier, M., and
		Robert}]{Martinetal2018}
	Martin, G.~M., McCabe, B.~P.~M., Frazier, D.~T., M., W., Robert, C.~P.:
	Auxiliary Likelihood-Based Approximate Bayesian Computation in State Space
	Models. J. Comput. Graph. Stat. 0\penalty0 (0), \penalty0 1--31 (2019)
	
	\bibitem[{Mattingly et~al.(2002)Mattingly, Stuart, and Higham}]{3}
	Mattingly, J.~C., Stuart, A.~M., Higham, D.~J.: Ergodicity for SDEs and
	approximations: locally Lipschitz vector fields and degenerate noise. Stoch.
	Proc. Appl. 101\penalty0 (2), \penalty0 185--232 (2002)
	
	\bibitem[{Maybank et~al.(2017)Maybank, Bojak, and Everitt}]{Maybank2017}
	Maybank, P., Bojak, I., Everitt, R.: Fast approximate Bayesian inference for
	stable differential equation models  (2017)
	\eprint{https://arxiv.org/abs/1706.00689}
	
	\bibitem[{McKinley et~al.(2017)McKinley, Vernon, Andrianakis, McCreesh, Oakley,
		Nsubuga, Goldstein, and White}]{35}
	McKinley, T.~J., Vernon, I., Andrianakis, I., McCreesh, N., Oakley, J.,
	Nsubuga, R., Goldstein, M., White, R.: Approximate Bayesian Computation and
	Simulation-Based Inference for Complex Stochastic Epidemic Models. Stat. Sci.
	33\penalty0 (1), \penalty0 4--18 (2017)
	
	\bibitem[{Mclachlan and Quispel(2002)}]{Mclachlan2002}
	Mclachlan, R., Quispel, G.: Splitting methods. Acta Numer. 11, \penalty0
	341--434 (2002)
	
	\bibitem[{Milstein and Tretyakov(2004)}]{5}
	Milstein, G.~N., Tretyakov, M.~V.: Stochastic numerics for mathematical
	physics. Scientific computation: Springer, Berlin (2004)
	
	\bibitem[{Misawa(2001)}]{REF_I7}
	Misawa, T.: A Lie Algebraic Approach to Numerical Integration of Stochastic
	Differential Equations. SIAM J. Sci. Comput. 23\penalty0 (3), \penalty0
	866--890 (2001)
	
	\bibitem[{Moores et~al.(2015)Moores, Drovandi, Mengersen, and
		Robert}]{Mooresetal2015}
	Moores, M.~T., Drovandi, C.~C., Mengersen, K., Robert, C.~P.: Pre-processing
	for approximate Bayesian computation in image analysis. Stat. Comput. 25,
	\penalty0 23--33 (2015)
	
	\bibitem[{Mori et~al.(2016)Mori, Mediburu, and Lozano}]{REF_I4}
	Mori, U., Mediburu, A., Lozano, J.~A.: Distance measures for time series in R:
	The TSdist package. R. J. 8, \penalty0 455--463 (2016)
	
	\bibitem[{Moro and Schurz(2007)}]{REF13_2}
	Moro, E., Schurz, H.: Boundary Preserving Semianalytic Numerical Algorithms for
	Stochastic Differential Equations. SIAM J. Sci. Comput. 29, \penalty0
	1525--1549 (2007)
	
	\bibitem[{Muskulus and Verduyn-Lunel(2011)}]{11}
	Muskulus, M., Verduyn-Lunel, S.: Wasserstein distances in the analysis of time
	series and dynamical systems. Physica. D. 240\penalty0 (1), \penalty0 45--58
	(2011)
	
	\bibitem[{Picchini(2014)}]{43}
	Picchini, U.: Inference for SDE models via Approximate Bayesian Computation. J.
	Comput. Graph. Stat. 23\penalty0 (4), \penalty0 1080--1100 (2014)
	
	\bibitem[{Picchini and Forman(2016)}]{44}
	Picchini, U., Forman, J.~L.: Accelerating inference for diffusions observed
	with measurement error and large sample sizes using Approximate Bayesian
	Computation. J. Stat. Comput. Simul. 86\penalty0 (1), \penalty0 195--213
	(2016)
	
	\bibitem[{Picchini and Samson(2018)}]{53}
	Picchini, U., Samson, A.: Coupling stochastic EM and approximate Bayesian
	computation for parameter inference in state-space models. Comput. Stat.
	33\penalty0 (1), \penalty0 179--212 (2018)
	
	\bibitem[{Pons(2011)}]{REF_I2}
	Pons, O.: Functional Estimation for Density, Regression Models and Processes.
	World Scientific Publishing, Singapore (2011)
	
	\bibitem[{Prangle(2017)}]{Prangle2017}
	Prangle, D.: Adapting the abc distance function. Bayesian Anal. 12\penalty0
	(1), \penalty0 289--309 (2017)
	
	\bibitem[{Prangle(2018)}]{32}
	Prangle, D.: Summary Statistics in Approximate Bayesian Computation. In:
	Handbook of Approximate Bayesian Computation: Chapman {\&} Hall: pp 125--152
	(2018)
	
	\bibitem[{Prangle et~al.(2014)Prangle, Blum, Popovic, and Sisson}]{28}
	Prangle, D., Blum, M.~G.~B., Popovic, G., Sisson, S.~A.: Diagnostic tools for
	approximate Bayesian computation using the coverage property. Aust. NZ. J.
	Stat. 56\penalty0 (4), \penalty0 309--329 (2014)
	
	\bibitem[{Quinn et~al.(2014)Quinn, Clarkson, and Mckilliam}]{REF_I5}
	Quinn, B., Clarkson, I., Mckilliam, R.: On the periodogram estimators of
	periods from interleaved sparse, noisy timing data. In: 2014 IEEE Stat.
	Signal Processing Workshop: pp 232--235 (2014)
	
	\bibitem[{{R Development Core Team}(2011)}]{R}
	{R Development Core Team}: R: A Language and Environment for Statistical
	Computing. R Foundation for Statistical Computing, Vienna, Austria (2011)
	
	\bibitem[{Robert(2016)}]{26}
	Robert, C.~P.: Approximate Bayesian Computation: A Survey on Recent Results.
	In: Cools, R., Nuyens, D. (eds) Monte Carlo and Quasi-Monte Carlo Methods:
	Springer International Publishing, Cham: pp 185--205 (2016)
	
	\bibitem[{van Rotterdam et~al.(1982)van Rotterdam, Lopes~da Silva, van~den
		Ende, Viergever, and Hermans}]{vanRotterdam1982}
	van Rotterdam, A., Lopes~da Silva, F., van~den Ende, J., Viergever, M.~A.,
	Hermans, A.: A model of the spatial-temporal characteristics of the alpha
	rhythm. Bull. Math. Biol. 44, \penalty0 283--305 (1982)
	
	\bibitem[{Sason and Verd\'{u}(2016)}]{REF_I6}
	Sason, I., Verd\'{u}, S.: $f$-Divergence Inequalities. IEEE T. Inform. Theory
	62\penalty0 (11), \penalty0 5973--6006 (2016)
	
	\bibitem[{Sisson et~al.(2018)Sisson, Fan, and Beaumont}]{sisson2018handbook}
	Sisson, S.~A., Fan, Y., Beaumont, M.: Handbook of Approximate Bayesian
	Computation. Chapman \& Hall/CRC Handbooks of Modern Statistical Methods: CRC
	Press, Taylor \& Francis Group (2018)
	
	\bibitem[{Str{\o}mmen~Melb{\o} and Higham(2004)}]{57}
	Str{\o}mmen~Melb{\o}, A.~H., Higham, D.~J.: Numerical simulation of a linear
	stochastic oscillator with additive noise. Appl. Numer. Math. 51\penalty0
	(1), \penalty0 89--99 (2004)
	
	\bibitem[{Sun et~al.(2015)Sun, Lee, and Hoeting}]{46}
	Sun, L., Lee, C., Hoeting, J.~A.: Parameter inference and model selection in
	deterministic and stochastic dynamical models via approximate Bayesian
	computation: modeling a wildlife epidemic. Environmetrics 26\penalty0 (7),
	\penalty0 451--462 (2015)
	
	\bibitem[{Tancredi(2019)}]{Tancredi}
	Tancredi, A.: Approximate Bayesian inference for discretely observed
	continuous-time multi-state models. Biometrics  (2019)
	\eprint{https://doi.org/10.1111/biom.13019}
	
	\bibitem[{Toni et~al.(2009)Toni, Welch, Strelkowa, Ipsen, and Stumpf}]{51}
	Toni, T., Welch, D., Strelkowa, N., Ipsen, A., Stumpf, M.~P.~H.: Approximate
	Bayesian computation scheme for parameter inference and model selection in
	dynamical systems. J. Roy. Soc. Interface 6\penalty0 (31), \penalty0 187--202
	(2009)
	
	\bibitem[{Vo et~al.(2015)Vo, Drovandi, Pettitt, and Simpson}]{50}
	Vo, B.~N., Drovandi, C.~C., Pettitt, A.~N., Simpson, M.~J.: Quantifying
	uncertainty in parameter estimates for stochastic models of collective cell
	spreading using approximate Bayesian computation. Math. Biosci. 263,
	\penalty0 133--142 (2015)
	
	\bibitem[{Wendling et~al.(2000)Wendling, Bellanger, Bartolomei, and
		Chauvel}]{60}
	Wendling, F., Bellanger, J.~J., Bartolomei, F., Chauvel, P.: Relevance of
	nonlinear lumped-parameter models in the analysis of depth-EEG epileptic
	signals. Biol. Cybern. 83\penalty0 (4), \penalty0 367--378 (2000)
	
	\bibitem[{Wendling et~al.(2002)Wendling, Bartolomei, Bellanger, and
		Chauvel}]{59}
	Wendling, F., Bartolomei, F., Bellanger, J.~J., Chauvel, P.: Epileptic fast
	activity can be explained by a model of impaired GABAergic dendritic
	inhibition. Eur. J. Neurosci. 15\penalty0 (9), \penalty0 89--99 (2002)
	
	\bibitem[{Wood(2010)}]{Wood2010}
	Wood, S.~N.: Statistical inference for noisy nonlinear ecological dynamic
	systems. Nature 466\penalty0 (7310), \penalty0 1102 (2010)
	
	\bibitem[{Zhu et~al.(2016)Zhu, Marin, and Leisen}]{47}
	Zhu, W., Marin, J.~M., Leisen, F.: A Bootstrap Likelihood Approach to Bayesian
	Computation. Aust. NZ. J. Stat. 58\penalty0 (2), \penalty0 227--244 (2016)
	
\end{thebibliography}

\end{document}